\documentclass{aa}  
\usepackage{graphicx}
\usepackage{txfonts}
\usepackage{amsmath}
\usepackage{physics}
\usepackage{amsfonts}
\usepackage{amssymb}
\usepackage{makecell}
\usepackage{tabularx}
\usepackage{xcolor}

\begin{document}

   \title{A semi-analytical model of the outer structure of protoplanetary discs formed by the collapse of a rotating molecular cloud}

   %\subtitle{I. Overviewing the $\kappa$-mechanism}

   \author{A. Anyiszonyan
          \inst{1}
          \and
          Zs. Sándor\inst{1}\inst{2}\inst{3}
          }

   \institute{ELTE Eötvös Loránd University, Institute of Physics and Astronomy, Department of Astronomy\\
              Pázmány Péter sétány 1/A, H-1117 Budapest, Hungary\\
              \email{A.Anyiszonyan@astro.elte.hu; Zs.Sandor@astro.elte.hu}
         \and
             Konkoly Observatory, HUN-REN Research Centre for Astronomy and Earth Sciences,\\ Konkoly Thege Mikl\'os \'ut 15-17., 1121 Budapest, Hungary
         \and
             CSFK, MTA Centre of Excellence, Konkoly-Thege Miklós út 15-17., 1112 Budapest, Hungary\\
             }

   \titlerunning{Outer structure of protoplanetary discs}
   \date{Received; accepted}

% \abstract{}{}{}{}{} 
% 5 {} token are mandatory
 
  \abstract
  % context heading (optional)
  % {} leave it empty if necessary  
   { Protoplanetary discs are formed due to the fragmentation and collapse of giant molecular cloud cores. The physical properties and structure of a formed disc are of great importance when studying the onset of planet formation processes.}
  % aims heading (mandatory)
   { Starting from the isothermal collapse of a rotating Bonnor-Ebert sphere, and assuming the conservation of angular momentum, we look for the structure equations of the newly formed protoplanetary disc. We take into account the possible role of pressure gradient in forming the initial disc structure, and compare our results with those obtained from a ``Keplerian'' infall model. Our aim is to obtain initial conditions to numerically study the evolution of the gaseous and solid components of protoplanetary discs.}
  % methods heading (mandatory)
   { The structure equations developed for protoplanetary discs have been derived analytically, while these equations have been solved numerically.}
  % results heading (mandatory)
   { The surface density profiles of the newly formed protoplanetary discs strongly depend on the initial rotation state of the Bonnor-Ebert sphere. According to our results, for slow rotators, gravitational instabilities can develop in the early phases of disc formation, while for relatively fast rotators, the outermost regions of the resulting discs are gravitationally stable, quite massive and highly sub-Keplerian, allowing rapid dust transport to the inner disc and subsequent planet formation.}
  % conclusions heading (optional), leave it empty if necessary 
   {}
  
   \keywords{ISM:clouds --
                Protoplanetary disks --
                Method:analytical
               }

    \maketitle
%
%-------------------------------------------------------------------

\section{Introduction}
Theoretical research on the formation of stars and their accretion discs is far from its conclusion. Even if one restricts the investigation to the formation of a single star from a molecular gas cloud, the field is still widely open, since the various clouds differ from one another in many parameters (form, mass, motion, strength of magnetic field etc.), and their collapse can be triggered by several effects (mass accretion and/or slow compression leading to the infringement of the stability limit, a shock wave created by a supernova explosion etc.). During the collapse, the physical interactions, which play an essential role, vary in space and time. The description of the collapse is therefore a challenging task. 

The elaboration of the ``inside-out'' collapse model of isothermal gas clouds had begun with \cite{Shu1977}. The present-day version of the model reckons with several different phases of the collapse. (For a brief summary on the topic, see, e.g., \cite{Basu2015}.) The first one is a ``free fall'' phase, which is essentially isothermal (the cloud remains optically thin, and its cooling is provided by the radiation of molecular hydrogen and dust particles). This phase ends with the formation of an optically thick ``first core'' in the cloud's centre. Further mass accretion leads to the heating of the core. When the temperature has reached cca. 2000 K, a massive dissociation of hydrogen molecules begins, and the core starts to collapse again. The second collapse is halted by the ionization of hydrogen atoms and the formation of the protostar, which continues to accrete material from the surrounding disc.

The cloud core collapse has also been modeled as a successive infall of thin spherical shells onto the disc, during which each gas parcel is accreted at a radius where its specific angular momentum matches the local Keplerian one \citep[see][for example]{HuesoGuillot2005A&A, Takahashi2013}. In the work of \cite{HuesoGuillot2005A&A}, the outer part of the disc is fed by the infalling material from the remnant of the collapsing molecular cloud, and the governing equation of the evolution of the (axisymmetric) disc \citep{Lynden-BellPringle1974MNRAS} is augmented by a source term responsible for the infalling material. 

In the work of \cite{Takahashi2013}, already during the infall of the thin spherical shells, the disc's angular momentum is promptly redistributed by the gravitational torques that are awaken for the Toomre parameter \citep{Toomre1964ApJ} falling in the interval $Q < 2$  by using an effective $\alpha$ viscosity parametrized by $Q$. The effect of the disc's radial pressure gradient on its rotation and mass distribution, which, however, according to our present study might be important, is neglected in their model.

To study the time evolution of the collapse of a molecular cloud, when a proto-disc is already formed around the protostar might be relevant, since under some specific circumstances, planetesimal formation can already begin, when the protoplanetary disc has not yet formed entirely \citep{DrazkowskaDullemond2018A&A}. In this phase, the outer part of the disc is fed by the infalling material from the remnant of the collapsing molecular cloud. 

For the formation of star-planet systems, the most important processes take place in the inner, densest region of the collapsing cloud. As we can see, its evolution is a highly complicated process. Those less dense parts of the cloud that are further away from the rotation axis will eventually form the outer part of the disc. From the viewpoint of the formation of the star-planet system, the main role of this outer part is to gradually absorb the excess angular momentum from the disc's inner part, thus making accretion possible by promoting the outward transport of angular momentum. Since it also contains solid material, the outer disc might be able to supply material for planet formation for a long enough time interval in the form of inward-directed pebble flux, being necessary for the formation of larger planetesimals by the streaming instability \citep{YoudinGoodman2005} and their further growth to planetary cores and protoplanets mainly via pebble accretion \citep{LambrechtsJohansen2012A&A}. To estimate the inward transport of the solid particles, one has to know the structure of the disc's outer part, where the gas dynamics may significantly deviate from Keplerian rotation due to a steep radial pressure gradient. Our aim is therefore to describe the state of the disc far from the centre, right after it has formed, i.e., in the epoch where the disc is already sufficiently ``flat'' to use a two-dimensional model, but the gas parcels still have the same angular momentum as before the collapse of the molecular cloud.

In this work, we derive the structure and motion of the disc's outer part from the state of the progenitor molecular cloud it has formed from. The latter is described as an isothermal gas sphere, i.e., a Bonnor-Ebert (BE) sphere (see \cite{Bonnor1956} and \cite{Ebert1955}). Despite its simplicity, this model can be also a good approximation of real cases \citep[cf.][]{Alves2001}. We assume that initially the BE sphere rotates as a rigid body, being slightly supercritical.

The most important result of our work is that for quickly rotating cloud cores, the pressure gradient may play a significant role in shaping the structure of the outer disc. To highlight the importance of the role played by the radial pressure gradient within the forming disc, we compare our results with those based on the model of \cite{Takahashi2013}.
   
%--------------------------------------------------------------------
\section{Description of the physical model}
\label{sec:Description_of_the_physical_model}
We suppose a non-magnetized interstellar gas cloud with a homogeneous chemical composition and zero temperature gradient, which is slowly rotating as a solid body. The slow rotation implies that the cloud’s mass distribution is essentially spherically symmetric, i.e., it is a Bonnor-Ebert sphere. However, if the ratio of central density to density on the edge surpasses a critical value of about $14.1$, the system is unstable, and a small perturbation leads to its gravitational collapse. This may eventually lead to the formation of a central protostar and a thin accretion disc around it. 

We aim to model the final state of the collapsing BE sphere without following the dynamics of the collapse. This can be achieved by using some assumptions. First, we suppose that the cloud retains its axial symmetry throughout the collapse. We also suppose that the temperature remains constant during the collapse. This assumption will surely not hold for the central, denser parts of the collapsing cloud, but can be accepted for the parts farther away from the axis of rotation, which will eventually form the outer part of the disc. As a further assumption, we neglect the transport of angular momentum during the collapse, at least for the above-mentioned outer parts of the cloud. Finally, we also neglect the mass loss of the cloud, which can take place if the inner part loses its angular momentum by emitting bipolar jets. These basic assumptions enable us to set up equations, which connect the structure of the final state to that of the initial state, from which the collapse begins. The solution of these equations gives the surface density and angular velocity profiles of the resulting accretion disc.

Throughout the text, we'll refer to the outer region of the disc where the assumptions described in the previous paragraph hold as the ``outer'' disc.

\section{Basic equations of structure}
\label{sec:Basic_equation_of_structure}

\subsection{The initial state}

As the initial state of the gas cloud, we assume a rotating Bonnor-Ebert sphere that can be characterized by the following five parameters: (i) the total cloud mass $M_\mathrm{BE}$, (ii) the mean molecular weight $\mu$, (iii) the temperature $T$, (iv) the ratio of rotational to gravitational energies $\beta$, and (v) the ratio of densities in the centre $\varrho_0$ and at the edge of the gas sphere 
\begin{equation}
    K= \frac{\varrho_0}{\varrho(R_\mathrm{BE})},
\end{equation}
where $R_\mathrm{BE}$ is the radius of the Bonnor-Ebert sphere.

The density distribution of the initial Bonnor-Ebert sphere can be obtained by solving the Emden-Chandrasekhar equation (see \cite{Emden1907}, \cite{Chandrasekhar1939}):
\begin{equation}
\label{emden_chandra}
\frac{1}{\xi^2} \frac{\dd}{\dd\xi}\left(\xi^2\frac{\dd\psi}{\dd\xi}\right)=e^{-\psi},
\end{equation}
where
\begin{equation}
\psi = \ln \left(\frac{\varrho_0}{\varrho(r)}\right)
\end{equation}
is the natural logarithm of the density ratio in the centre and at a certain radius, $r$, and the dimensionless coordinate $\xi$ is related to the physical one by
\begin{equation}
\xi=k \cdot r
\end{equation}
with the constant
\begin{equation}
\label{eqn:k}
k=\sqrt{\frac{4\pi G \varrho_0\mu m_\mathrm{p}}{k_\mathrm{B} T}}.
\end{equation}
Analogously, $\xi_\mathrm{max}$ is the value of $\xi$ on the edge of the sphere:
\begin{equation}
\label{eqn:k-def}
\xi_\mathrm{max}=k \cdot R_\mathrm{BE}.
\end{equation}
The value of $\xi_\mathrm{max}$ is determined by the density ratio $K$. To calculate it, one must either integrate the Emden-Chandrasekhar equation numerically or use an approximate solution. In our numerical calculations, we have used a formula given by \cite{Iacono2014}, which is highly accurate for the relevant values of $\xi$. Thus, the ratio of density at $\xi$ to the central density, denoted here by $F(\xi)$, can be expressed as
\begin{equation}
\label{eqn:F_xi-function}
F(\xi)=\left(1+\frac{\xi^2}{6}+\frac{\xi^4}{180}\cdot\frac{1+4\xi^2/189}{1+\xi^2/27+\xi^4/4115}\right)^{-1}.
\end{equation}
For a given value of $K$, the value of $\xi_\mathrm{max}$ can be calculated by solving the algebraic equation
\begin{equation}
K=\frac{1}{F(\xi_\mathrm{max})}=1+\frac{\xi_\mathrm{max}^2}{6}+\frac{\xi_\mathrm{max}^4}{180}\cdot\frac{1+4\xi_\mathrm{max}^2/189}{1+\xi_\mathrm{max}^2/27+\xi_\mathrm{max}^4/4115}.
\end{equation}
The value of $k$ is then given by
\begin{equation}
\label{eqn:k-2}
k=\frac{k_\mathrm{B}T}{GM_\mathrm{BE}\mu m_\mathrm{p}} \cdot \int_0^{\xi_\mathrm{max}} \xi^2 F(\xi) \dd{\xi}.
\end{equation}
(The detailed derivation of the above equation can be found in Appendix A.)

\subsection{Conservation of angular momentum}
Assuming the conservation of the angular momentum of each infalling gas parcel, one can obtain one of the equations which describe the structure of the accretion disc. According to \cite{Machida2007}, for the regions far away from the axis of rotation, angular momentum can be conserved if there is no magnetic field and axial symmetry breaking. (The presence of a magnetic field can lead to angular momentum transport via magnetic braking, while the lack of axial symmetry results in gravitational torques.)

We assume that the disc's angular velocity $\Omega$ and surface density $\Sigma$ only depend on the radial coordinate $r$. To characterize the initial state, we introduce $I(j)$ which is the semi-column density of the BE sphere parametrized by the specific angular momentum $j=\tilde r^2\omega$ at a given distance $\tilde r$ from the axis of rotation, where $\omega$ denotes the angular velocity of the rigidly rotating Bonnor-Ebert sphere. Thus  $I(j)$ can be written as
\begin{equation}
\label{eqn:I_j-function}
I(j)=\int_0^{\sqrt{R_{\mathrm{BE}}^2-\frac{j}{\omega}}} \varrho\left(\sqrt{\frac{j}{\omega}+z^2}\right) \dd{z}.
\end{equation}
The conservation of the angular momentum of each gas parcel falling from the cloud onto the final circular orbit in the disc can be expressed by the following equation:
\begin{equation}
\label{eqn:dO_dr}
\frac{\dd{\Omega}}{\dd{r}}=\frac{\omega}{r \cdot I(j)} \cdot \Sigma - \frac{2\Omega}{r}
\end{equation}
(See the details in Appendix B.)

Figure \ref{fig:infall-scheme} presents a schematic illustration of the model used.Figure 1 presents a schematic illustration of the model used.

\begin{figure}[h]
\caption{A schematic illustration of the model used. Each gas parcel of the BE sphere whose specific angular momentum falls into a given interval will eventually be accreted onto a ring within the disc, while retaining its angular momentum. As a result of a radial pressure gradient, the specific angular momentum of the gas within the ring can differ from the Keplerian value. The force vectors drawn on the figure give rise to a sub-Keplerian rotation.}
\label{fig:infall-scheme}
\centering
\includegraphics[width=1\columnwidth]{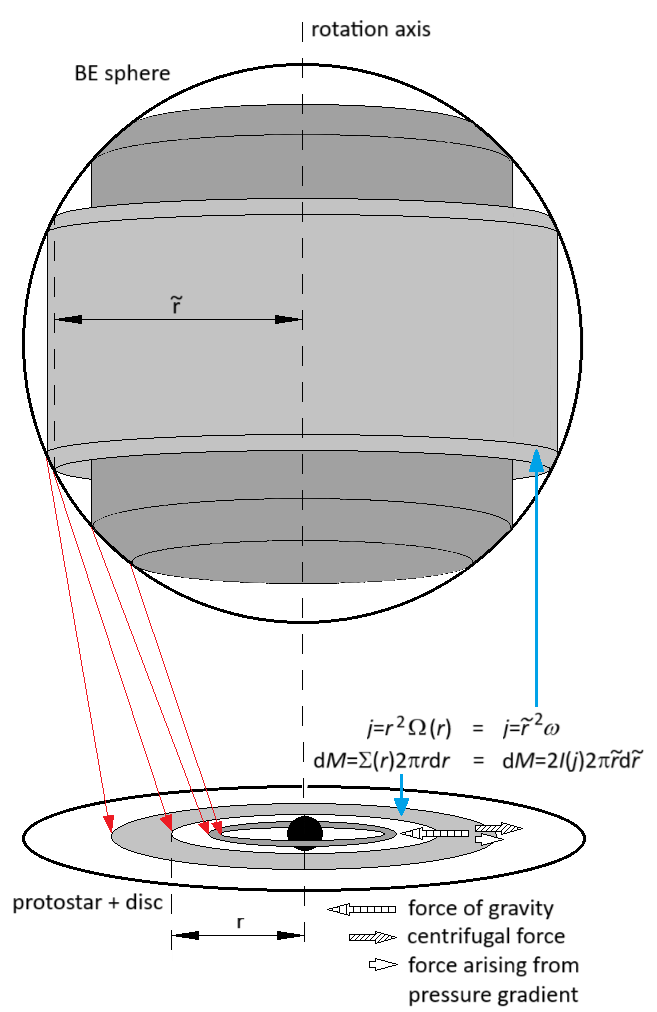}
\end{figure}

\subsection{Balance of forces}
Assuming a circular motion for each gas parcel, the balance between centrifugal force, gravity and pressure gradient yields another equation for the structure of the disc. To derive the latter, we start from the Euler equation:
\begin{equation}
\varrho \left(\frac{\partial \mathbf{v}}{\partial t} + \left( \mathbf{v} \nabla \right) \mathbf{v} \right) = - \nabla p - \varrho \nabla U,
\end{equation}
where $U$ stands for the gravitational potential. Using cylindrical coordinates, for the radial components we have:
\begin{equation}
- \frac{\varrho v^2}{r}  = - \frac{\partial p}{\partial r} - \varrho \frac{\partial U}{\partial r}
\end{equation}

Substituting $\Omega$ for $\frac{v}{r}$ and $c^2_{\mathrm{s}}\varrho$ for $p$, where $c_{\mathrm{s}}$ stands for the isothermal sound speed (which, in our case, is independent of $r$), we can write:
\begin{equation}
- \varrho r \Omega^2  = - c^2_{\mathrm{s}} \frac{\partial \varrho}{\partial r} - \varrho \frac{\partial U}{\partial r},
\end{equation}
which can be integrated with respect to $z$. Rearranging this equation, we get the following:
\begin{equation}
\label{eqn:dS_dr}
\frac{\dd{\Sigma}}{\dd{r}}=\frac{1}{c_{\mathrm{s}}^2}\left(r\cdot\Omega^2-\frac{\dd{U}}{\dd{r}}\right)\Sigma.
\end{equation}
\textbf
(Let us note that we have used the thin disc approximation, supposing that within the disc, $\frac{\partial U}{\partial r}$ does not depend on $z$.)

\subsection{The complete set of equations}
The set of equations \eqref{eqn:I_j-function}, \eqref{eqn:dO_dr} and \eqref{eqn:dS_dr} must be complemented with an equation for the gravitational potential $U$ and another one for the specific angular momentum. The latter one is:
\begin{equation}
j=r^2 \cdot \Omega.
\end{equation}

To set up an equation for the gravitational potential, some other assumptions are needed, since gravitational forces involve significant contributions from the inner and more massive parts of the resulting configuration, for which our basic assumptions do not hold. We shall return to this topic in subsection \ref{grav_pot_gradient}.
 
\section{Transformed equations of structure}
\label{sec:Transformed_equations_of_structure}

\subsection{Introducing new functions}
The basic equations as they are presented in the previous section are difficult to handle, therefore it is appropriate to transform the functions involved in such a way that makes the equations easier to solve. This will be done in this section. For further explanations, see Appendix C.

One has to set an outer boundary for the disc (otherwise equation (\ref{eqn:dS_dr}) would imply an infinite total mass). In the following, we will use the same outer boundary for the disc as for the initial Bonnor-Ebert sphere, i.e., the maximum value for the radial coordinate $r$ will be equal to $R_{\mathrm{BE}}$. (The more general case will be examined in subsection \ref{different_boundary}.) Because of angular momentum conservation, $\Omega(R_{\mathrm{BE}})$ equals $\omega$, i.e., the initial angular velocity of the Bonnor-Ebert sphere.

We introduce the dimensionless spatial coordinate $x$, the value of which is 0 at $r=R_{\mathrm{BE}}$ and 1 at $r=0$:
\begin{equation}
x=1-\frac{r}{R_{\mathrm{BE}}}.
\end{equation}
At $r=R_{\mathrm{BE}}$, the value of $I(j(r))$ is zero while the surface density $\Sigma$ is not, which makes the derivative $\frac{\dd{\Omega}}{\dd{r}}$ tend to infinity as $r$ approaches to $R_{\mathrm{BE}}$ (cf. equation (\ref{eqn:dO_dr})). Therefore, we introduce a new function $\chi(x)$, which has a finite (but non-zero) derivative at $x=0$:
\begin{equation}
\label{def:chi}
\chi(x)=\left(1-\frac{j(r(x))}{j_{\mathrm{max}}}\right)^{\frac{3}{2}}=\left[1-\frac{(1-x)^2}{\omega}\Omega(R_{\mathrm{BE}}(1-x))\right]^{\frac{3}{2}},
\end{equation}
where $j_{\mathrm{max}}$ denotes the maximum specific angular momentum:
\begin{equation}
j_{\mathrm{max}}=R_{\mathrm{BE}}^2 \cdot \omega.
\end{equation}
Instead of $\Sigma(r)$, $I(j)$ and $U(r)$ we introduce $\sigma(x)$, $Y(\chi)$ and $V(x)$, respectively:
\begin{equation}
\sigma(x)=\ln\frac{\Sigma(r(x))}{\Sigma(R_{\mathrm{BE}})}=\ln\frac{\Sigma(R_{\mathrm{BE}}(1-x))}{\Sigma(R_{\mathrm{BE}})},
\end{equation}
\begin{equation}
Y(\chi)=I(j(\chi))=I\left(j_{\mathrm{max}}\left(1-\chi^{2/3}\right)\right),
\end{equation}
\begin{equation}
V(x)=U(r(x))=U(R_{\mathrm{BE}}(1-x)).
\end{equation}
Instead of equations (\ref{eqn:dO_dr}) and (\ref{eqn:dS_dr}) we now have:
\begin{equation}
\label{eqn:dchi_dx}
\frac{\dd \chi}{\dd x} = \frac{3}{2}\Sigma(R_{\mathrm{BE}})\frac{\chi^{1/3}}{Y(\chi)}(1-x)e^{\sigma(x)},
\end{equation}
\begin{equation}
\label{eqn:ds_dx}
\frac{\dd \sigma}{\dd x} = - \frac{R_{\mathrm{BE}}^2\omega^2}{c_{\mathrm{s}}^2}\frac{\left(1-\chi^{2/3}\right)^2}{(1-x)^3} - \frac{1}{c_{\mathrm{s}}^2}\frac{\dd V}{\dd x},
\end{equation}
with the following initial conditions:
\begin{equation}
\label{eqn:initial_conditions}
\left\{
\begin{aligned}
\chi(0)=0,\\
\sigma(0)=0.
\end{aligned} \right.
\end{equation}

\subsection{The gradient of gravitational potential}
\label{grav_pot_gradient}
Since the surface density is expected to rise quickly from the edge of the disc towards the centre, it is reasonable to approximate the gradient of the gravitational potential at a given distance $r$ from the centre as if the potential gradient were produced by a point mass in the centre, which has a mass equal to the total mass within radius $r$. (In reality, rings with radii greater than $r$ exert an outward gravitational force on the gas at $r$. On the other hand, rings with smaller radii exert an inward force, which is greater than the gravitational force exerted by an equivalent point mass in the centre. For the gravitational potential of thin rings, see \cite{Lass1983}.)

The conservation of angular momentum (for the outer part of the disc) enables the calculation of the mass within a given radius $r$ from the value of $\chi$ at $r$. Introducing the function
\begin{equation}
\label{def:zeta}
\zeta(\chi)=\frac{\int_0^{1-\chi^{2/3}}\left(\int_0^{\sqrt{1-\varepsilon}} F\left(\xi_{\mathrm{max}} \cdot \sqrt{\varepsilon + z^2} \right) \dd z \right) \dd \varepsilon}{\int_0^1\left(\int_0^{\sqrt{1-\varepsilon}} F\left(\xi_{\mathrm{max}} \cdot \sqrt{\varepsilon + z^2} \right) \dd z \right) \dd \varepsilon},
\end{equation}
which gives the ratio of mass within $r(\chi)$ to the total mass, the gradient $\frac{\dd V}{\dd x}$ can be approximated as
\begin{equation}
\label{eqn:dV_dx}
\frac{\dd V}{\dd x} = - \frac{GM_{\mathrm{BE}}\zeta(\chi)}{R_{\mathrm{BE}}(1-x)^2}.
\end{equation}

\begin{figure}[h]
\caption{Function $\zeta(\chi)$ for $K=15$}
\label{fig:zeta_chi-K15}
\centering
\includegraphics[width=0.8\columnwidth]{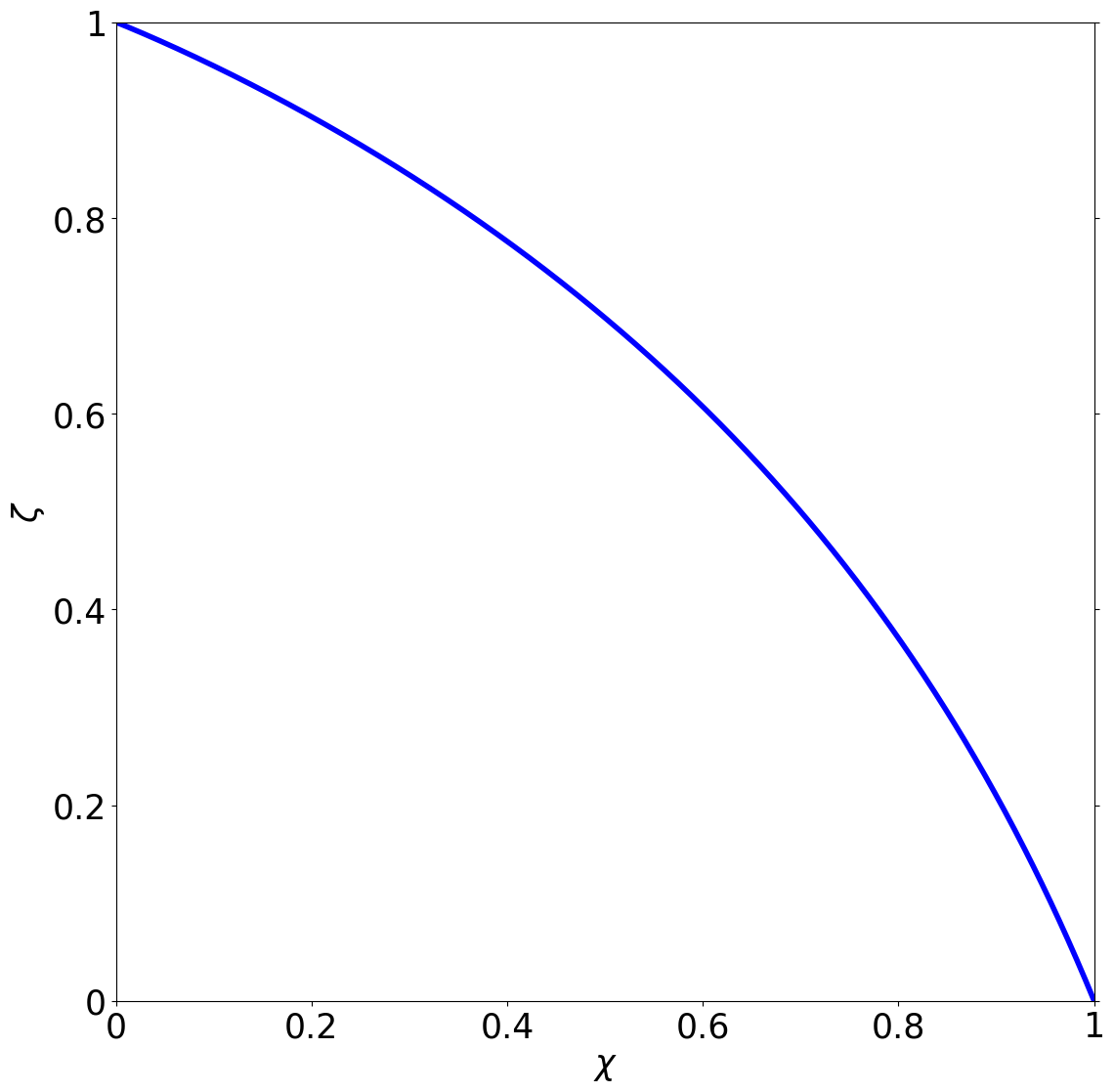}
\end{figure}

For the derivation of the expression for $\zeta(\chi)$ given by equation (\ref{def:zeta}), see Appendix D. The function $\zeta(\chi)$ for $K=15$ is illustrated in Figure \ref{fig:zeta_chi-K15}.

\subsection{The final form of the equations of structure}
We can insert the formula given by equation (\ref{eqn:dV_dx}) into equation (\ref{eqn:ds_dx}). To write down the resulting equation in a more succinct form, we introduce 3 dimensionless parameters (energy ratios), the value of each determined solely by $\xi_{\mathrm{max}}$ (or the density ratio $K$):
\begin{equation}
\kappa=\frac{E_{\mathrm{BE,rot}}}{M_{\mathrm{BE}} R_{\mathrm{BE}}^2 \omega^2}=\frac{\int_0^{\xi_{\mathrm{max}}}\xi^4 F(\xi) \dd \xi}{3\xi_{\mathrm{max}}^2 \cdot {\int_0^{\xi_{\mathrm{max}}}\xi^2 F(\xi) \dd \xi}},
\end{equation}
\begin{equation}
\lambda=\frac{|E_{\mathrm{BE,grav}}|}{\frac{GM_{\mathrm{BE}}^2}{R_{\mathrm{BE}}}}=\frac{\xi_{\mathrm{max}} \cdot \int_0^{\xi_{\mathrm{max}}} \left( \int_0^{\xi} \xi '^2 F(\xi ') \dd \xi ' \right) \xi F(\xi) \dd \xi}{\left( \int_0^{\xi_{\mathrm{max}}}\xi^2 F(\xi) \dd \xi \right)^2},
\end{equation}
\begin{equation}
\tau=\frac{|E_{\mathrm{BE,grav}}|}{\lambda M_{\mathrm{BE}} c_{\mathrm{s}}^2}=\frac{\int_0^{\xi_{\mathrm{max}}}\xi^2 F(\xi) \dd \xi}{\xi_{\mathrm{max}}}.
\end{equation}
The values of $\xi_{\mathrm{max}}$, $\kappa$, $\lambda$ and $\tau$ for some values of $K$ are given in Table \ref{table:parameters}.
\begin{table}[ht]
\caption{Values of $\xi_{\mathrm{max}}$, $\kappa$, $\lambda$ and $\tau$ for some values of $K$}
\centering
\begin{tabular}{c c c c}
\hline\hline
$K$ & 15 & 20 & 25 \\
\hline
$\xi_{\mathrm{max}}$ & 6.6272 & 7.4450 & 8.1395 \\
$\kappa$ & 0.1400 & 0.1344 & 0.1302 \\
$\lambda$ & 0.7368 & 0.7575 & 0.7745 \\
$\tau$ & 2.4481 & 2.4921 & 2.5107 \\
\hline
\end{tabular}
\label{table:parameters}
\end{table}
\noindent
Now equation (\ref{eqn:ds_dx}) can be written the following form:
\begin{equation}
\label{eqn:ds_dx-final}
\frac{\dd \sigma}{\dd x}= \tau \left( \frac{\zeta(\chi)}{(1-x)^2} - \frac{\beta \lambda (1-\chi^{2/3})^2}{\kappa (1-x)^3} \right),
\end{equation}
for the details of the calculations, see Appendix E.

\begin{figure}[h]
\caption{Function $W(\chi)$ for $K=15$}
\label{fig:W_chi-K15}
\centering
\includegraphics[width=1\columnwidth]{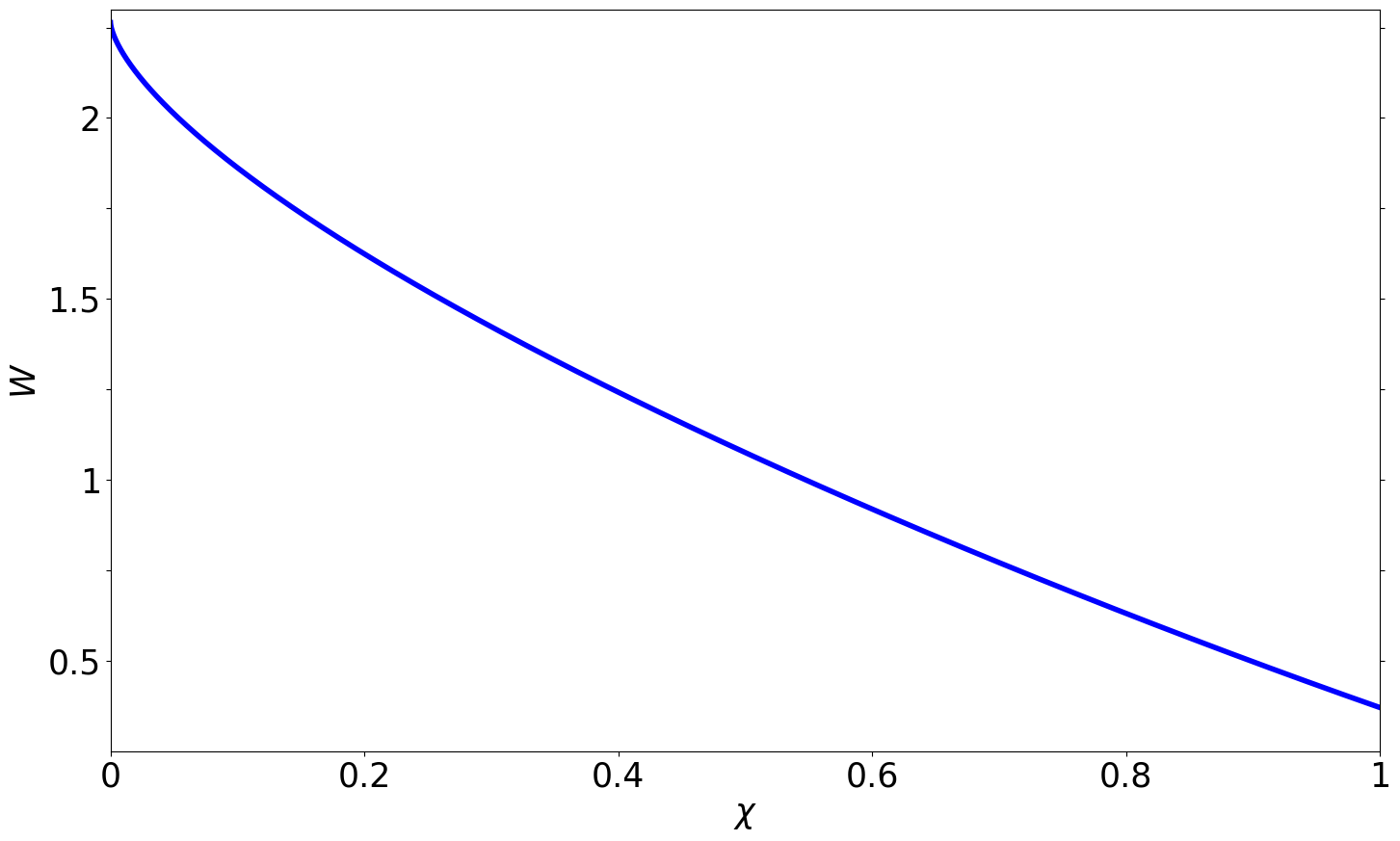}
\end{figure}

Equation (\ref{eqn:dchi_dx}) can also be written in a more succinct form. For this end, we introduce the dimensionless function $W(\chi)$ (determined solely by $\xi_{\mathrm{max}}$) and the constant $C$:
\begin{equation}
\label{eqn:W_chi}
W(\chi) = \frac{\varrho(0)}{k} \frac{\chi^{1/3}}{Y(\chi)},
\end{equation}
\begin{equation}
\label{eqn:C}
C = \frac{3k\Sigma(R_{\mathrm{BE}})}{2\varrho(0)}.
\end{equation}
One can easily check that
\begin{equation}
\lim_{\chi \rightarrow 0^+} W(\chi) = \frac{K}{\xi_{\mathrm{max}}}.
\end{equation}
Instead of Equation \eqref{eqn:dchi_dx}, we now have
\begin{equation}
\label{eqn:dchi_dx-final}
\frac{\dd \chi}{\dd x} = C \cdot W(\chi) (1-x) e^{\sigma(x)}.
\end{equation}
The function $W(\chi)$ for $K=15$ is displayed in Figure \ref{fig:W_chi-K15}.

\subsection{Method of solution}

As we have seen, the equations of structure for the ``outer'' disc (for which our basic assumptions hold) and the initial conditions can be written as follows:

\begin{equation}
\label{eqn:system_of_equations_of_structure}
\left\{
\begin{aligned}
\frac{\dd \sigma}{\dd x}= \tau \left( \frac{\zeta(\chi)}{(1-x)^2} - \frac{\beta \lambda (1-\chi^{2/3})^2}{\kappa (1-x)^3} \right),\\
\frac{\dd \chi}{\dd x} = C \cdot W(\chi) (1-x) e^{\sigma(x)},\\
\chi(0)=0,\\
\sigma(0)=0.
\end{aligned} \right.
\end{equation}

For each value of the positive parameter $C$ there's a single solution.

It is not possible to take a realistic guess for the value of $C$ \textit{a priori}. Instead, one can proceed by fixing at first the mass ratio of the ``outer'' disc to the whole system, this way obtaining the value of $\zeta$ at the inner edge of the ``outer'' disc. We will denote this value by $\zeta_{\mathrm{in}}$ (i.e., the mass ratio of the ``outer'' disc to the whole system is equal to $1-\zeta_{\mathrm{in}}$) and the corresponding value of $\chi$ by $\chi_{\mathrm{in}}$ (i.e., $\zeta(\chi_{\mathrm{in}})=\zeta_{\mathrm{in}}$). As a next step, one can fix the ratio of the local angular velocity to the Keplerian one at the inner edge. These two ratios determine   $C$ and $x_{\mathrm{in}}$, the latter denoting the value of the dimensionless coordinate at the inner edge (i.e., $\zeta(\chi(x_{\mathrm{in}}))=\zeta(\chi_{\mathrm{in}})=\zeta_{\mathrm{in}}$). By elevating both ratios simultaneously, we can obtain the same solution, but within an extended domain (with a higher $x_{\mathrm{in}}$). Thus, it is reasonable to suppose super-Keplerian rotation at the inner edge, i.e., to set the angular velocity ratio above unity at $x_{\mathrm{in}}$, in order to get a full picture of the characteristics of the possible solutions. In this way, we can easily identify the radial distance from the protostar, where the orbital motion of the gas becomes exactly Keplerian. Our model for the ``outer'' disc, therefore, can be used as initial condition for protoplanetary disc evolution after truncating the super-Keplerian region, and only considering the sub-Keplerian part.

In the following, we will use $\frac{\Omega}{\Omega_{\mathrm{Kep}}}=1.1$ at $x_{\mathrm{in}}$. Since the centrifugal force is proportional to $\Omega^2$, this means that the second term on the right-hand side of equation (\ref{eqn:ds_dx-final}) is equal to $1.1^2 = 1.21$ times the first term. It follows that

\begin{equation}
\label{eqn:x_in}
x_{\mathrm{in}}=1-\frac{\beta\lambda\left(1-\chi_{\mathrm{in}}^{2/3}\right)^2}{1.21\kappa\zeta_{\mathrm{in}}}.
\end{equation}

In order to obtain the actual value of the constant $C$, as well as the functions $\sigma(x)$ and $\chi(x)$, one can use successive approximations. That is, we integrate Equations \eqref{eqn:ds_dx-final} and \eqref{eqn:dchi_dx-final} from $x=0$ to $x=x_{\mathrm{in}}$ for different values of $C$ until we find a solution where $\chi(x_{\mathrm{in}})=\chi_{\mathrm{in}}$ holds with sufficiently high precision.

To solve \eqref{eqn:system_of_equations_of_structure}, we have written a computer code. For given values of $\mu$, $T$, $K$, $M_{\mathrm{BE}}$, $\beta$ and $\zeta_{\mathrm{in}}$, it first determines the values of $\xi_{\mathrm{max}}$, $\kappa$, $\lambda$, $\tau$, $\chi_{\mathrm{in}}$ and $x_{\mathrm{in}}$. After that, it starts to integrate the equations of structure from $x=0$, using an arbitrary value for $C$. The numerical integration is performed with a fourth-order Runge-Kutta method. If the value of $\chi$ exceeds the value of $\chi_{\mathrm{in}}$ before $x$ reaches $x_{\mathrm{in}}$, the integration is repeated with a smaller value for $C$. On the other hand, if the value of $\chi(x_{\mathrm{in}})$ stays below $\chi_{\mathrm{in}}$, the integration is repeated with a greater value for $C$. The integrations are repeated until the obtained value of $\chi(x_{\mathrm{in}})$ differs from $\chi_{\mathrm{in}}$ by less than 0.01\%. After that, the code transforms the functions $\chi(x)$ and $\sigma(x)$ into $\Omega(r)$ and $\Sigma(r)$. This code is available upon request.

\subsection{The question of outer boundary} \label{different_boundary}

In the previous treatment, we have chosen the outer boundary of the disc to be located at $R_{\mathrm{BE}}$. This choice is somewhat arbitrary, and it is important to check how sensitive the solutions are to the location of the boundary. Therefore we rewrite the equations of structure for the case where the outer boundary is placed at some arbitrary radius $R_{\mathrm{out}}$.
\noindent
First, we re-define the dimensionless coordinate $x$:
\begin{equation}
x=1-\frac{r}{R_{\mathrm{out}}},
\end{equation}
then we introduce the constant $\delta$ for the ratio of the outer radius to that of the Bonnor-Ebert sphere:
\begin{equation}
\delta=\frac{R_{\mathrm{out}}}{R_{\mathrm{BE}}}.
\end{equation}
The conservation of angular momentum yields an expression for $\Omega(R_{\mathrm{out}})$:
\begin{equation}
\Omega\left(R_{\mathrm{out}}\right) = \frac{j_{\mathrm{max}}}{R^2_{\mathrm{out}}} = \left(\frac{R_{\mathrm{BE}}}{R_{\mathrm{out}}}\right)^2\omega = \frac{\omega}{\delta^2}.
\end{equation}
\noindent
The functions $\chi(x)$ and $\sigma(x)$ will be the following:
\begin{equation}
\chi(x)=\left(1-\frac{j(r(x))}{j_{\mathrm{max}}}\right)^{\frac{3}{2}}=\left[1-\frac{\delta^2(1-x)^2}{\omega}\Omega(R_{\mathrm{out}}(1-x))\right]^{\frac{3}{2}},
\end{equation}
\begin{equation}
\sigma(x)=\ln\frac{\Sigma(r(x))}{\Sigma(R_{\mathrm{out}})}=\ln\frac{\Sigma(R_{\mathrm{out}}(1-x))}{\Sigma(R_{\mathrm{out}})}.
\end{equation}
One of the equations of the structure now reads:
\begin{equation}
\label{eqn:ds_dx-final-delta}
\frac{\dd \sigma}{\dd x}= \frac{\tau}{\delta} \left( \frac{\zeta(\chi)}{(1-x)^2} - \frac{\beta \lambda (1-\chi^{2/3})^2}{\kappa\delta (1-x)^3} \right).
\end{equation}
\noindent
Since the constant $C$ can be redefined as
\begin{equation}
C = \frac{3k\Sigma(R_{\mathrm{out}})\delta^2}{2\varrho(0)},
\end{equation}
the other equation of structure retains its previous form:
\begin{equation}
\label{eqn:dchi_dx-final-delta}
\frac{\dd \chi}{\dd x} = C \cdot W(\chi) (1-x) e^{\sigma(x)},
\end{equation}
while the initial conditions also remain the same:
\begin{equation}
\sigma(0)=\chi(0)=0.
\end{equation}
To get the solution, we can use the method of successive approximations described in the previous subsection. For $x_{\mathrm{in}}$ we now have:
\begin{equation}
x_{\mathrm{in}}=1-\frac{\beta\lambda\left(1-\chi_{\mathrm{in}}^{2/3}\right)^2}{1.21\kappa\delta\zeta_{\mathrm{in}}}.
\end{equation}
It is worth noting that the radius corresponding to $x_{\mathrm{in}}$ does not depend on the location of the outer boundary.

\section{Solutions}
\label{sec:Solutions}

In the following we will examine three groups of cases (denoted by A, B and C), each group containing four different subcases. We always suppose initially a Bonnor-Ebert sphere with the following characteristics:
\begin{equation}
\left\{
\begin{aligned}
\mu=2.3\\
T=20\,\textup{K}\\
K=15\\
M_{\mathrm{BE}}=1\,\textup{M}_{\odot}
\end{aligned} \right.
\end{equation}
\noindent   
(Let us remind that from the point of view of stability, the critical value of $K$ is about $14.1$, i.e., the Bonnor-Ebert sphere defined above is slightly supercritical.) The corresponding radius and central density of the Bonnor-Ebert sphere are:
\begin{equation}
\left\{
\begin{aligned}
R_{\mathrm{BE}}=5050\,\mathrm{AU}\\
\varrho(0)=6.587 \cdot 10^{-18}\,\frac{\mathrm{g}}{\mathrm{cm}^3}
\end{aligned} \right.
\end{equation}
The three groups of cases differ in angular velocity, i.e., they have different values for $\beta$. According to observational data (see \cite{Caselli2002}), $\beta$ is typically between $10^{-4}$ and $0.07$, so we have chosen values from this interval. In group A, the Bonnor-Ebert sphere rotates the slowest, with $\beta=0.0025$. For groups B and C, the value of $\beta$ is 0.01 and 0.03, respectively. Within each group, each case has a different value for $\zeta_{\mathrm{in}}$. The four cases correspond to $\zeta_{\mathrm{in}}$ equal to 0.99, 0.95, 0.85 and 0.75.

\subsection{Group A}
\begin{figure}[h]
\centering
\includegraphics[width=1\columnwidth]{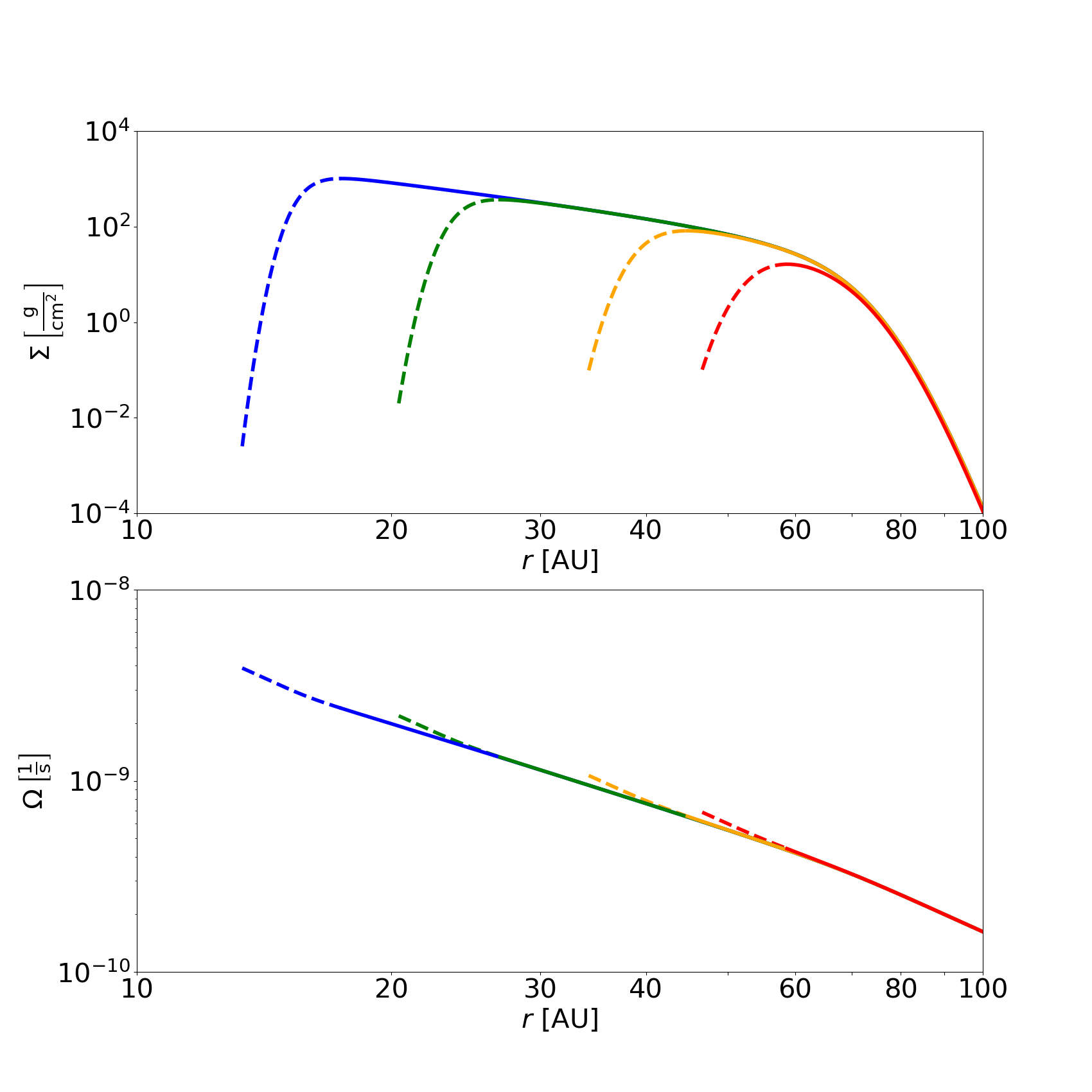}
\caption{Functions $\Sigma(r)$ and $\Omega(r)$ for $\beta=0.0025$ (Group A). Different colours correspond to different values of $\zeta_{\mathrm{in}}$. Red: $\zeta_{\mathrm{in}}=0.99$, orange: $\zeta_{\mathrm{in}}=0.95$, green: $\zeta_{\mathrm{in}}=0.85$, blue: $\zeta_{\mathrm{in}}=0.75$.}
\label{fig:Sigma_Omega_A}
\end{figure}

For the cases in this group, $\beta$ is equal to 0.0025. The resulting $\Sigma(r)$ and $\Omega(r)$ functions for the four different $\zeta_{\mathrm{in}}$ values are presented on Figure \ref{fig:Sigma_Omega_A}. One can see that the four different cases of group A yield practically the same $\Sigma(r)$ and $\Omega(r)$ functions, apart from the lower end of their domains.

\begin{figure}[h]
\centering
\includegraphics[width=1\columnwidth]{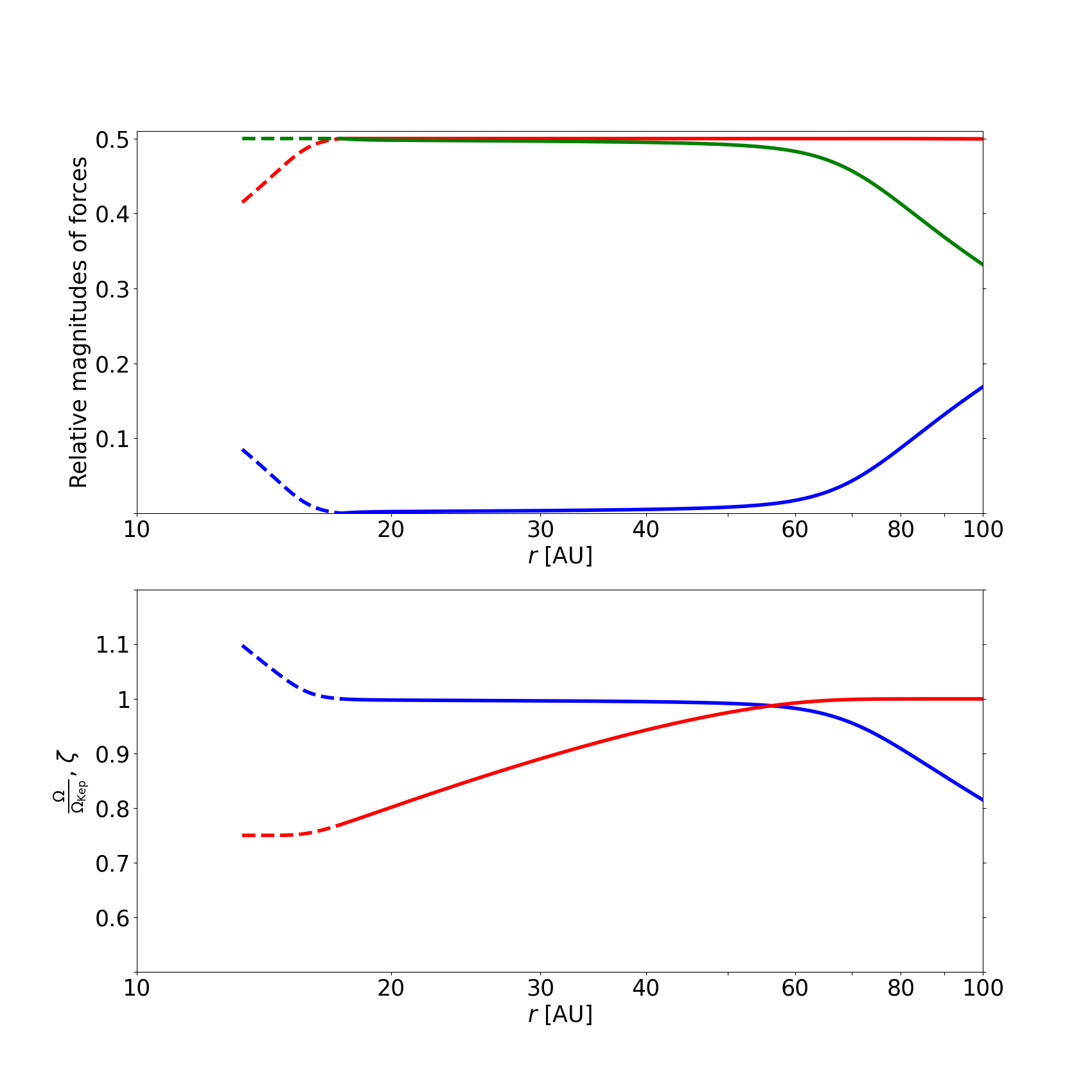}
\caption{Upper figure: The relative magnitudes of gravity (red), centrifugal force (green) and the force corresponding to pressure gradient (blue). Each magnitude is normalized to the sum of the three. \\ Lower figure: Functions $\zeta(r)$ (red) and $\frac{\Omega(r)}{\Omega_{\mathrm{Kep}}(r)}$ (blue). \\ Figures correspond to $\beta=0.0025$ (Group A) and $\zeta_{\mathrm{in}}=0.75$.}
\label{fig:zeta_Omega_per_OmegaK_forces_A}
\end{figure}

In each case, the ``outer'' disc can be divided into three parts or regions. On the log-log plot, the surface density profile increases with growing $r$ in the innermost region, decreases in the middle region, and drops with a much larger slope in the outermost region. (Very far from the centre, the slope becomes smaller again, but this region is out of interest because of its extremely small mass.) Passing from one region to another corresponds to a change in the relative magnitudes of the forces acting on the disc matter (cf. Figure \ref{fig:zeta_Omega_per_OmegaK_forces_A}, upper figure, for the case with $\zeta_{\mathrm{in}}=0.75$).

For small values of $r$, where rotation is super-Keplerian, $\Sigma(r)$ rises quickly with growing $r$. (Let us remind, however, that by setting up the condition expressed by eq. (\ref{eqn:x_in}), we artificially ``enforced'' the emergence of this super-Keplerian region. Hence, in each figure, the super-Keplerian region is presented with a dashed line, highlighting the fact that this region must be truncated in order to fit the ``outer'' disc to a Keplerian ``inner'' disc model.) In this region, both gravity and the force arising from pressure gradient are directed inwards, together counterbalancing the centrifugal force. For the case with $\zeta_{\mathrm{in}}=0.75$, the super-Keplerian region extends from $13.33\,\mathrm{AU}$ to $17.48\,\mathrm{AU}$ (where $\Omega$ becomes equal to $\Omega_{\mathrm{Kep}}$), and its mass is about $2\%$ of $M_{\mathrm{BE}}$ (or $8\%$ of the ``outer'' disc's total mass). At $17.48\,\mathrm{AU}$, the surface density reaches its maximum value of $1014\frac{\mathrm{g}}{\mathrm{cm}^2}$.

The second part of the ``outer'' disc is characterized by slightly sub-Keplerian rotation. Within this region, surface density diminishes with growing distance from the centre, approximately following a power law (with an exponent between $-2$ and $-3$). Centrifugal force and gravity are almost equal in magnitude, pressure gradient plays an insignificant role in establishing a balance of forces. For $\zeta_{\mathrm{in}}=0.75$, the ratio of $\Omega$ to $\Omega_{\mathrm{Kep}}$ diminishes slowly from unity at $17.48\,\mathrm{AU}$, reaching $0.99$ at $53.15\,\mathrm{AU}$. We choose the latter value of $r$ (somewhat arbitrarily) as the upper boundary of the ``outer'' disc's second part. This is the region where the bulk of the mass is concentrated: it comprises about $21\%$ of $M_{\mathrm{BE}}$ (or $84\%$ of the ``outer'' disc's mass). (Cf. the $\zeta(r)$ function on the lower figure of Figure \ref{fig:zeta_Omega_per_OmegaK_forces_A}: The largest change in the value of $\zeta$ occurs in the ``slightly sub-Keplerian'' region.)

For even larger values of $r$ (third region), the rotation of the disc becomes highly sub-Keplerian. The $\Sigma(r)$ function drops much quicker than before, the force arising from pressure gradient becomes important in counterbalancing gravity. For $\zeta_{\mathrm{in}}=0.75$, the third region contains $2\%$ of $M_{\mathrm{BE}}$ ($8\%$ of the ``outer'' disc's mass). Essentially, the disc vanishes around $90\,\mathrm{AU}$: beyond $90.59\,\mathrm{AU}$, $\zeta$ approaches unity closer than $10^{-6}$, i.e., the mass beyond that radius is only a millionth part of $M_{\mathrm{BE}}$.

\subsection{Group B}
\begin{figure}[h]
\centering
\includegraphics[width=1\columnwidth]{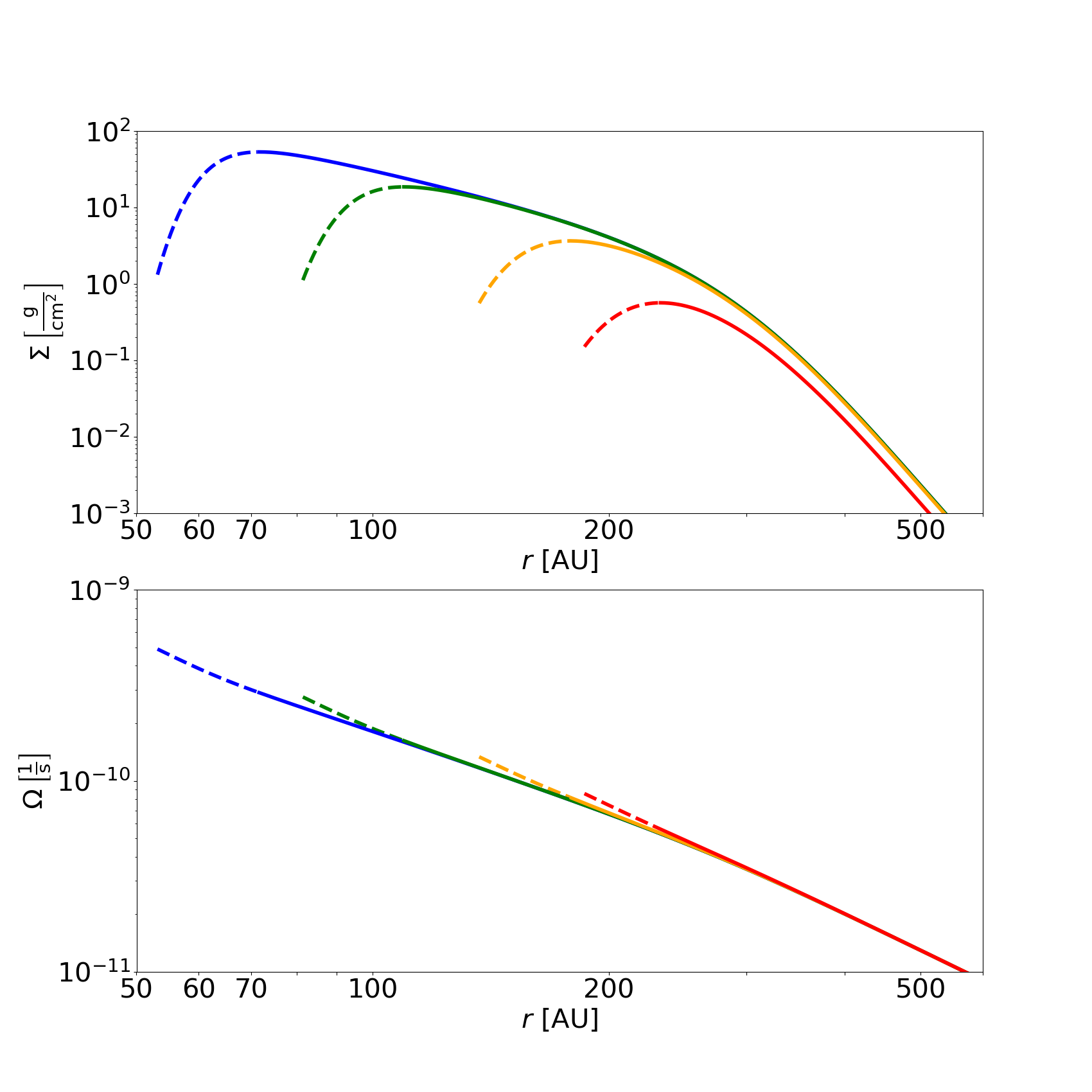}
\caption{Functions $\Sigma(r)$ and $\Omega(r)$ for $\beta=0.01$ (Group B). Different colours correspond to different values of $\zeta_{\mathrm{in}}$. Red: $\zeta_{\mathrm{in}}=0.99$, orange: $\zeta_{\mathrm{in}}=0.95$, green: $\zeta_{\mathrm{in}}=0.85$, blue: $\zeta_{\mathrm{in}}=0.75$.}
\label{fig:Sigma_Omega_B}
\end{figure}

\begin{figure}[h]
\centering
\includegraphics[width=1\columnwidth]{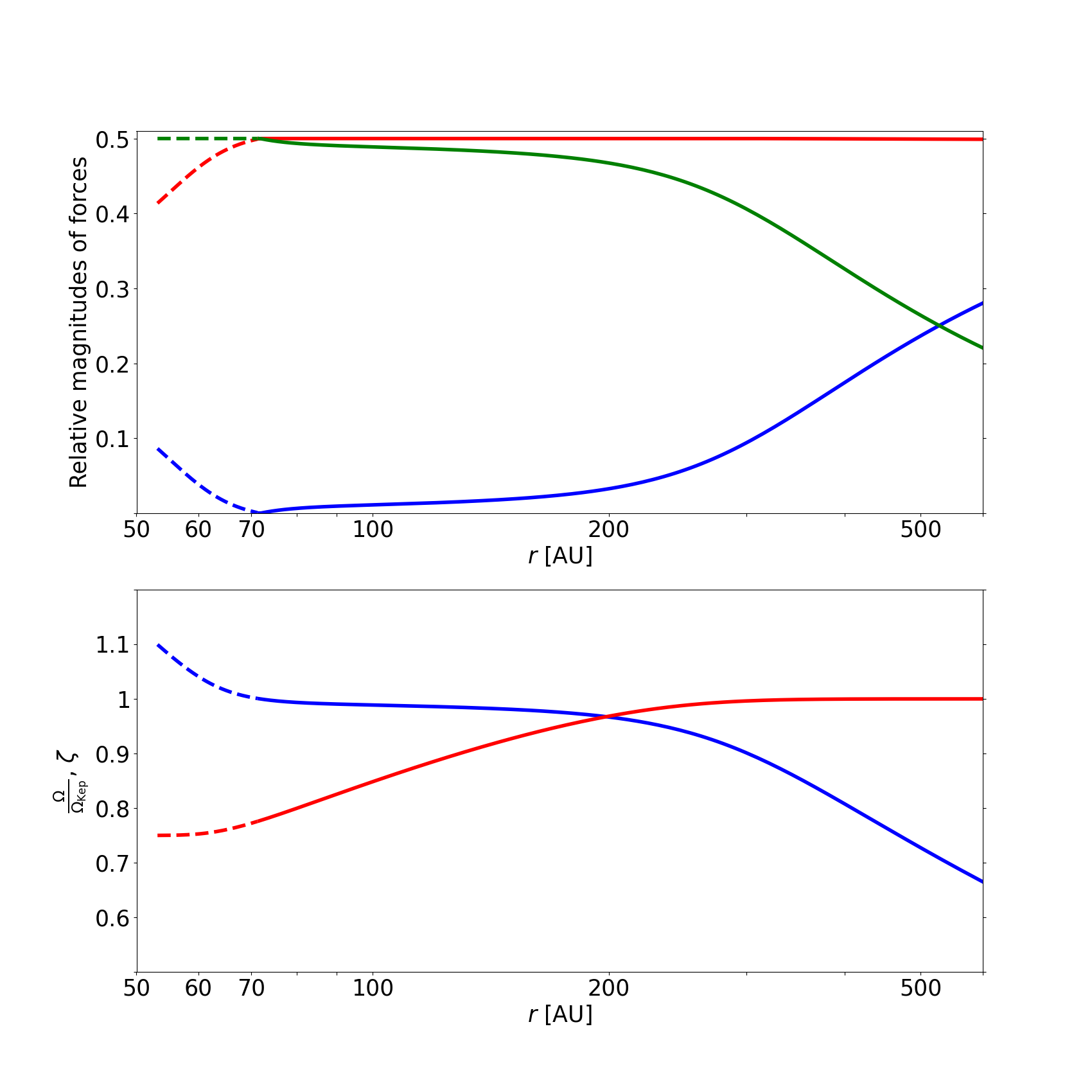}
\caption{Upper figure: The relative magnitudes of gravity (red), centrifugal force (green) and the force corresponding to pressure gradient (blue). Each magnitude is normalized to the sum of the three. \\ Lower figure: Functions $\zeta(r)$ (red) and $\frac{\Omega(r)}{\Omega_{\mathrm{Kep}}(r)}$ (blue). \\ Figures correspond to $\beta=0.01$ (Group B) and $\zeta_{\mathrm{in}}=0.75$.}
\label{fig:zeta_Omega_per_OmegaK_forces_B}
\end{figure}

For the cases in group B, $\beta$ equals 0.01. The resulting $\Sigma(r)$ and $\Omega(r)$ functions  are presented on Figure \ref{fig:Sigma_Omega_B}. Figure \ref{fig:zeta_Omega_per_OmegaK_forces_B} illustrates the relative magnitudes of forces as well as $\zeta$ and $\frac{\Omega}{\Omega_{\mathrm{Kep}}}$ as functions of the radial coordinate in the case with $\zeta_{\mathrm{in}}=0.75$.

The surface density profiles for $\zeta_{\mathrm{in}}=0.95$, 0.85 and 0.75 essentially converge to each other for large values of $r$, while the $\Sigma(r)$ function for $\zeta_{\mathrm{in}}=0.99$ stays somewhat below the other three. Apart from the latter case, one can still distinguish the three regions of the ``outer'' disc, but the passage from one region to another is much smoother than in the cases belonging to Group A. As for the $\Omega(r)$ functions in the four cases, they all still converge to each other at large radii.

In the case with $\zeta_{\mathrm{in}}=0.75$, the first (super-Keplerian) region is located between $53.18\,\mathrm{AU}$ and $71.86\,\mathrm{AU}$, its mass being about $2.7\%$ of $M_{\mathrm{BE}}$ or $11\%$ of the ``outer'' disc's mass. After a less dramatic rise than in the corresponding case of Group A, the surface density reaches its peak value of $53.3\,\frac{\mathrm{g}}{\mathrm{cm}^2}$. In the second (``slightly sub-Keplerian'') region, $\Sigma(r)$ follows a power law with a similar exponent as in the corresponding case of Group A. However, this time the second region is narrower and much less heavy. The angular velocity drops to $99\%$ of the local Keplerian one already at $92.53\,\mathrm{AU}$, and only $5.4\%$ of $M_{\mathrm{BE}}$ ($22\%$ of the ``outer'' disc's total mass) is located between $71.86$ and $92.53\,\mathrm{AU}$. The rest ($16.9\%$ of $M_{\mathrm{BE}}$, $67\%$ of the ``outer'' disc's total mass) is in the third (``highly sub-Keplerian'') part of the ``outer'' disc. The third region, in contrast to the second one, is much more extended than in the corresponding case of Group A: $\zeta$ approaches unity closer than $10^{-6}$ only beyond $766.7\,\mathrm{AU}$.

\subsection{Group C}
\begin{figure}[h]
\centering
\includegraphics[width=1\columnwidth]{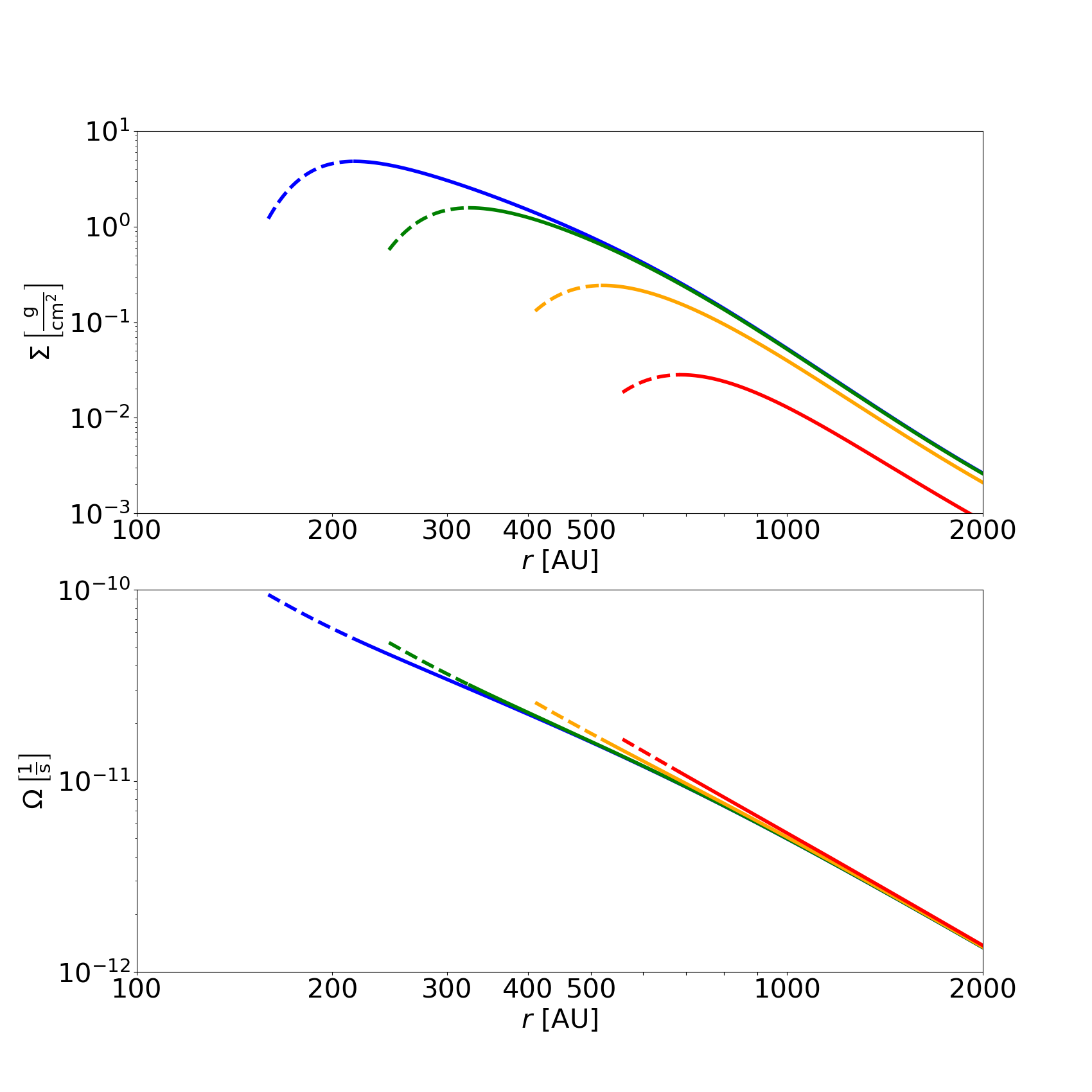}
\caption{Functions $\Sigma(r)$ and $\Omega(r)$ for $\beta=0.03$ (Group C). Different colours correspond to different values of $\zeta_{\mathrm{in}}$. Red: $\zeta_{\mathrm{in}}=0.99$, orange: $\zeta_{\mathrm{in}}=0.95$, green: $\zeta_{\mathrm{in}}=0.85$, blue: $\zeta_{\mathrm{in}}=0.75$.}
\label{fig:Sigma_Omega_C}
\end{figure}

\begin{figure}[h]
\centering
\includegraphics[width=1\columnwidth]{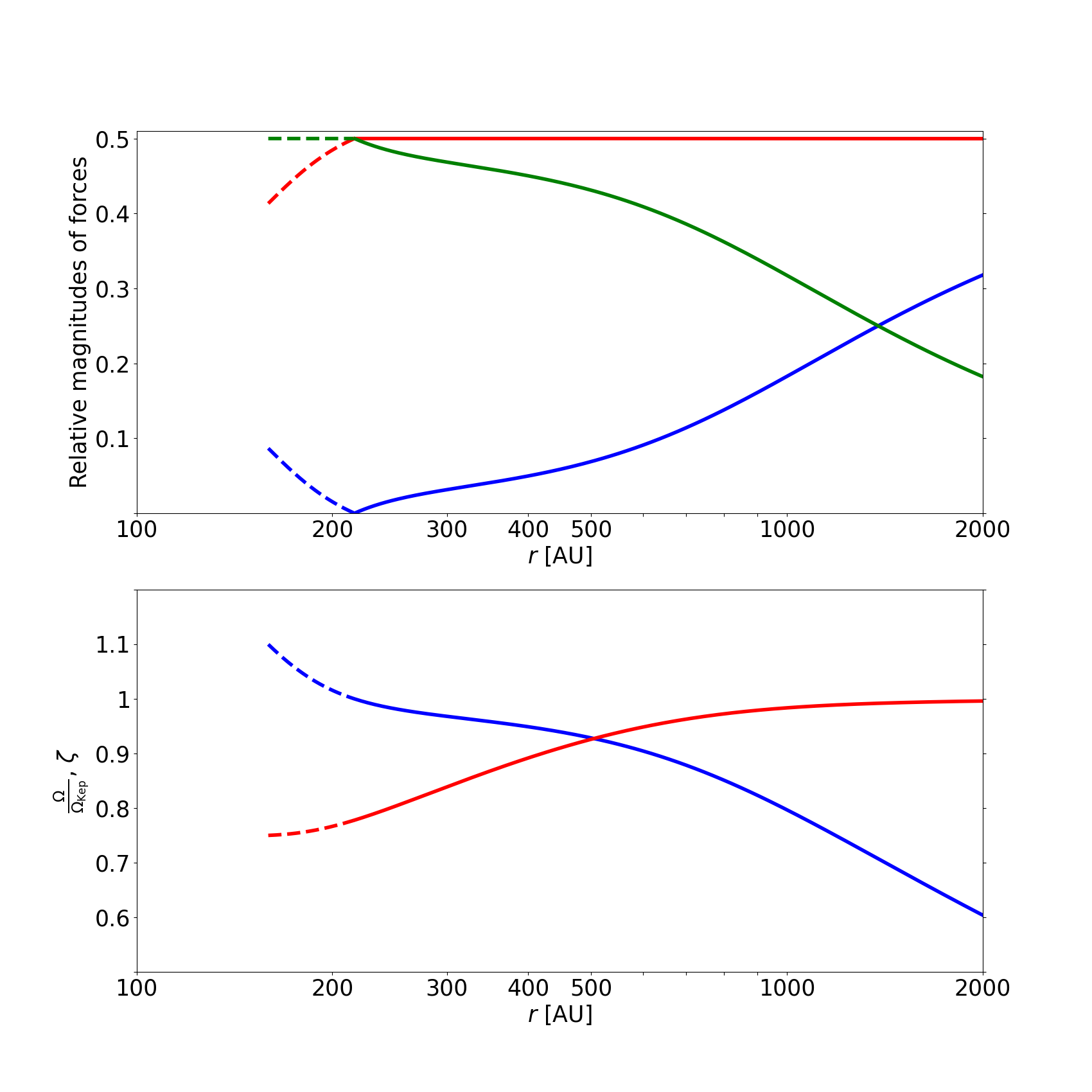}
\caption{Upper figure: The relative magnitudes of gravity (red), centrifugal force (green) and the force corresponding to pressure gradient (blue). Each magnitude is normalized to the sum of the three. \\ Lower figure: Functions $\zeta(r)$ (red) and $\frac{\Omega(r)}{\Omega_{\mathrm{Kep}}(r)}$ (blue). \\ Figures correspond to $\beta=0.03$ (Group C) and $\zeta_{\mathrm{in}}=0.75$.}
\label{fig:zeta_Omega_per_OmegaK_forces_C}
\end{figure}

The fastest initial rotation corresponds to the cases in group C, with $\beta$ equal to 0.03. The resulting $\Sigma(r)$ and $\Omega(r)$ functions  are presented on Figure \ref{fig:Sigma_Omega_C}. For the relative magnitudes of forces and the $\zeta(r)$ and $\frac{\Omega(r)}{\Omega_{\mathrm{Kep}}(r)}$ functions in the case with $\zeta_{\mathrm{in}}=0.75$, see Figure \ref{fig:zeta_Omega_per_OmegaK_forces_C}.

Here, the differences between the $\Sigma(r)$ profiles corresponding to different values of $\zeta_{\mathrm{in}}$ are even more remarkable than in Group B. From the three regions of the ``outer'' disc, the ``slightly sub-Keplerian'' one has almost completely disappeared: it is even narrower and lighter than in the cases of Group B. However, the qualitative behaviour of the $\Omega(r)$ functions is the same as in Groups A and B.

For $\zeta_{\mathrm{in}}=0.75$, the super-Keplerian region extends from $159.4\,\mathrm{AU}$ to $216.3\,\mathrm{AU}$, reaching at the end a peak surface density of $4.82\,\frac{\mathrm{g}}{\mathrm{cm}^2}$. It contains $2.8\%$ of $M_{\mathrm{BE}}$ (or $11.2\%$ of the ``outer'' disc's mass). The ``slightly sub-Keplerian'' region, between $216.3\,\mathrm{AU}$ and $231.8\,\mathrm{AU}$, now contains only $1.2\%$ of $M_{\mathrm{BE}}$ or $4.7\%$ of the ``outer'' disc's mass. The ``highly sub-Keplerian'' region extends from $231.8\,\mathrm{AU}$ virtually to the outer boundary of the disc at $5050\,\mathrm{AU}$, its mass being $21\%$ of $M_{\mathrm{BE}}$ (over $84\%$ of the ``outer'' disc's mass).

\subsection{Qualitative comparison of the three groups of cases}

It is hardly surprising that small values of $\beta$ yield a small disc. The example of Group A has shown that for a small $\beta$, the structure of the ``outer'' part of disc is almost insensitive to the structure of the ``inner'' part: whether the ``outer'' disc contains $1\%$ or $25\%$ of the disc matter, its ``tail'' has the same surface density profile and rotation curve. If the ``outer'' disc's mass is not negligibly small, the bulk of this mass moves around the centre with a slightly sub-Keplerian velocity, while the velocity of a small fraction on the ``edge'' is highly sub-Keplerian. (Of course, the small super-Keplerian part can be absent altogether, since each solution can be truncated ``from within''.)

With growing values of $\beta$, the size of the disc rises fast. The surface density profile of the ``outer'' part becomes more sensible to the structure of the ``inner'' part (but the same cannot be said about the rotation curve at large radii). The ``highly sub-Keplerian'' region grows together with $\beta$ at the expense of the ``slightly sub-Keplerian'' region, which shrinks both in mass and width (not just compared to the whole disc, but also in absolute terms).

\subsection{The question of stability}

The solutions obtained do not necessarily represent realistic cases which can persist for long, since gravitational instability may produce changes in the disc structure. According to Toomre's criterion (see \cite{Toomre1964ApJ}), instability occurs if $Q \lessapprox 1$, where the dimensionless parameter $Q$ can be calculated as follows:
\begin{equation}
Q = \frac{c_{\mathrm{s}} \kappa_\mathrm{e}}{\pi G \Sigma} \, ,
\end{equation}
where $\kappa_\mathrm{e}$ stands for the epicyclic frequency:
\begin{equation}
\kappa_\mathrm{e}^2 = \frac{2 \Omega}{r} \frac{\dd}{\dd  r} \left(r^2 \Omega \right)
\end{equation}
We have calculated the value of $Q$ as a function of the radial coordinate for all the cases examined above. In Group A, only the case with $\zeta_{\mathrm{in}}=0.99$ is stable on its whole domain. In Group B, there are three fully stable cases (albeit $Q$ gets close to unity in the case with $\zeta_{\mathrm{in}}=0.85$), while in Group C, due to the lower surface densities, all four cases are stable everywhere. The details are presented in Table \ref{table:stability}.

\begin{table}[ht]
\caption{Gravitational stability of the disc in the examined cases}
\centering
\begin{tabularx}{\columnwidth}{c X X X}
\hline\hline
 & A & B & C \\
\hline
$\zeta_{\mathrm{in}}=0.99$ & stable \mbox{everywhere} & stable \mbox{everywhere} & stable \mbox{everywhere} \\
$\zeta_{\mathrm{in}}=0.95$ & unstable \mbox{between} \mbox{$43$ and $47 \, \mathrm{AU}$} & stable \mbox{everywhere} & stable \mbox{everywhere} \\
$\zeta_{\mathrm{in}}=0.85$ & unstable \mbox{between} \mbox{$23$ and $49 \, \mathrm{AU}$} & stable \mbox{everywhere} & stable \mbox{everywhere} \\
$\zeta_{\mathrm{in}}=0.75$ & unstable \mbox{between} \mbox{$15$ and $49 \, \mathrm{AU}$} & unstable \mbox{between} \mbox{$62$ and $119 \, \mathrm{AU}$} & stable \mbox{everywhere} \\
\hline
\end{tabularx}
\label{table:stability}
\end{table}

If the initial assumptions that the collapse is isothermal and angular momentum is being conserved hold for those gas parcels as well, which eventually settle within the unstable region, then at a certain point axial symmetry will break, that may lead eventually to binary or giant planet formation (cf. \citep{Boss1997Sci}). Moreover, the resulting gravitational torques will affect the angular momentum transport, and thus the overall disc structure, as well. Hence, the discs in Group B and, especially, Group C seem more self-consistent, than those in Group A. In addition, one may expect that the initial assumptions hold better for the former groups, due to the larger distances from the centre as well as smaller densities and angular velocity gradients.

\subsection{Changing the outer boundary}

In the solutions we have obtained, the bulk of the mass and angular momentum is located within a radius far smaller than $R_{\mathrm{BE}}$. This suggests that we can displace the outer boundary considerably without seriously affecting the solutions: the effect of the displacement is virtually equivalent to a truncation or an extension of the curves of $\Sigma(r)$ and $\Omega(r)$ obtained for $R_{\mathrm{out}}=R_{\mathrm{BE}}$.

Numerical solutions of equations (\ref{eqn:ds_dx-final-delta}) and (\ref{eqn:dchi_dx-final-delta}) confirm these expectations. We will illustrate this in two examples, taken from Groups A and C. From both groups we choose the cases with $\zeta_{\mathrm{in}}=0.75$.

\begin{figure}[h]
\centering
\includegraphics[width=1\columnwidth]{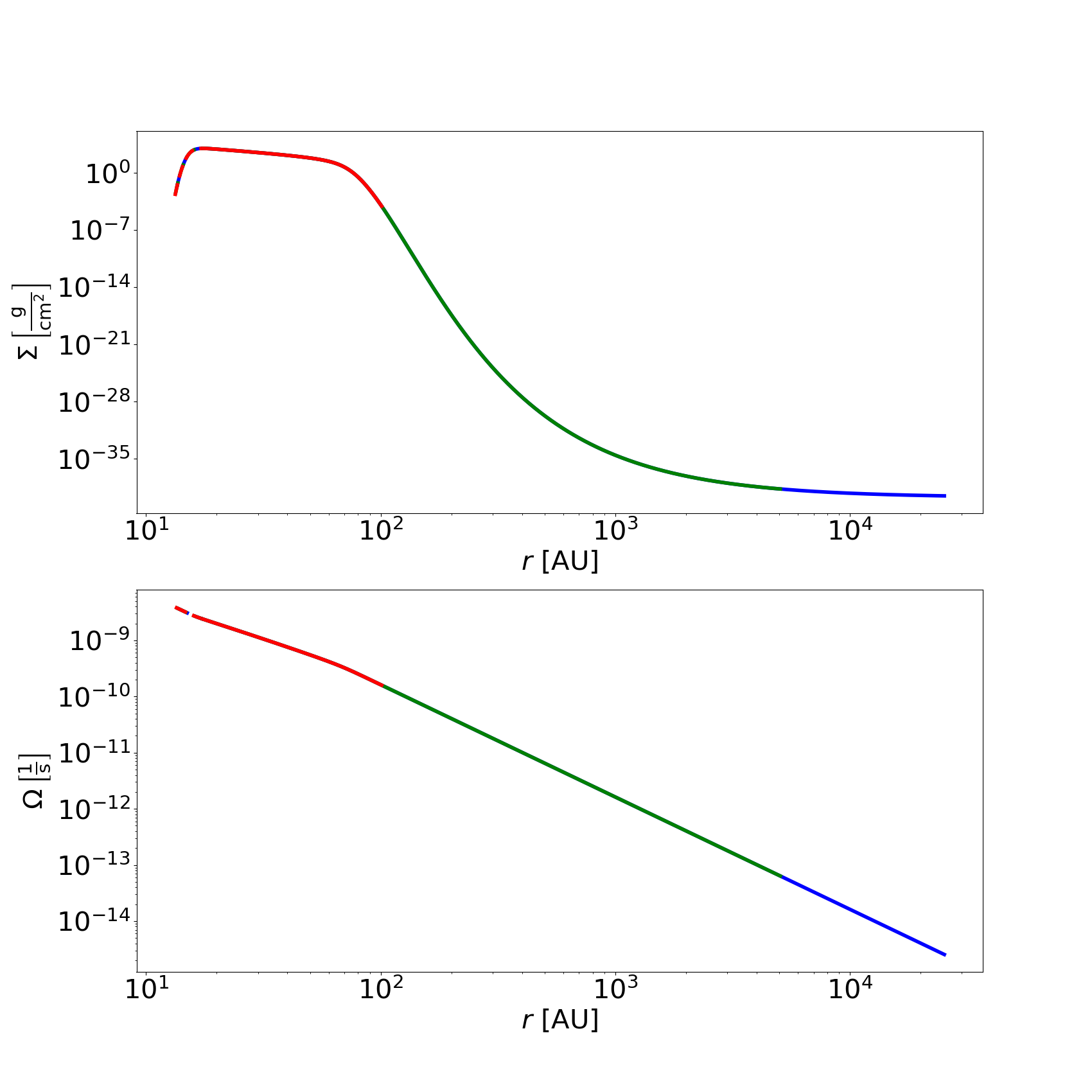}
\includegraphics[width=1\columnwidth]{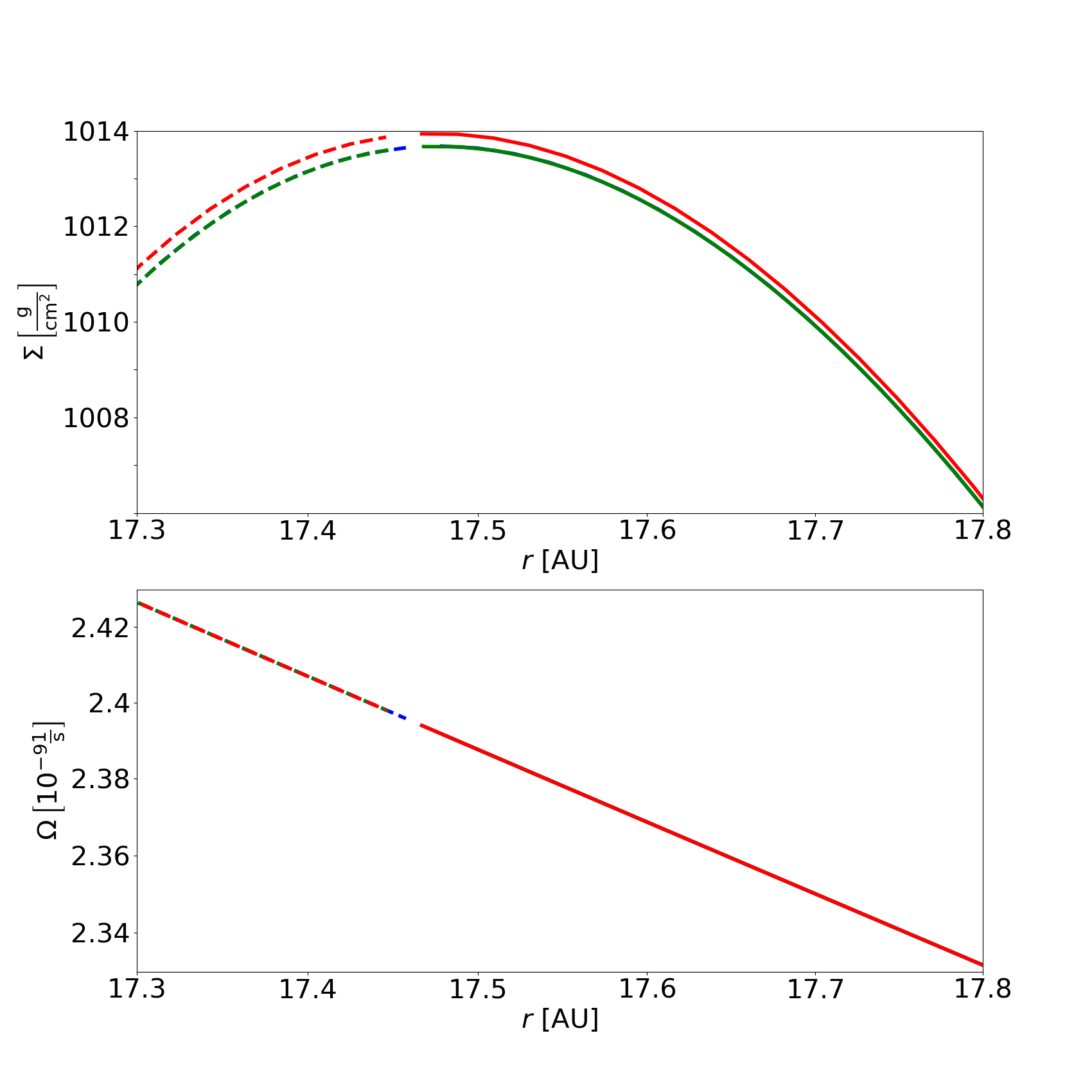}
\caption{Upper panel: Functions $\Sigma(r)$ and $\Omega(r)$ for $\beta=0.0025$ (Group A) and $\zeta_{\mathrm{in}}=0.75$ for outer boundaries located at 25,250 AU (blue), 5050 AU (green) and 101 AU (red). Lower panel: Enlarged images of the same functions around the maximum point of $\Sigma(r)$.}
\label{fig:A-mod}
\end{figure}

The upper panel of figure \ref{fig:A-mod} shows the solutions for $\beta=0.0025$ with three largely different outer boundaries, corresponding to $\delta=5$, $\delta=1$ and $\delta=0.02$ (cf. subsection \ref{different_boundary}). The outer boundaries are located thus at $R_{\mathrm{out}}=25,250 \, \mathrm{AU}$, $R_{\mathrm{out}}=5050 \, \mathrm{AU}$ and $R_{\mathrm{out}}=101 \, \mathrm{AU}$, respectively. (The inner boundary is at the same radius of $13.3\,\mathrm{AU}$ in each case.)  As one can see, at the given resolution of the figure, the curves overlap, i.e., the three solutions match each other almost perfectly. The lower panel of the same figure, which presents the same curves in the close vicinity of the maximum point of $\Sigma(r)$, confirms that the differences between the solutions obtained for different $\delta$ values are indeed extremely small.

\begin{figure}[h]
\centering
\includegraphics[width=1\columnwidth]{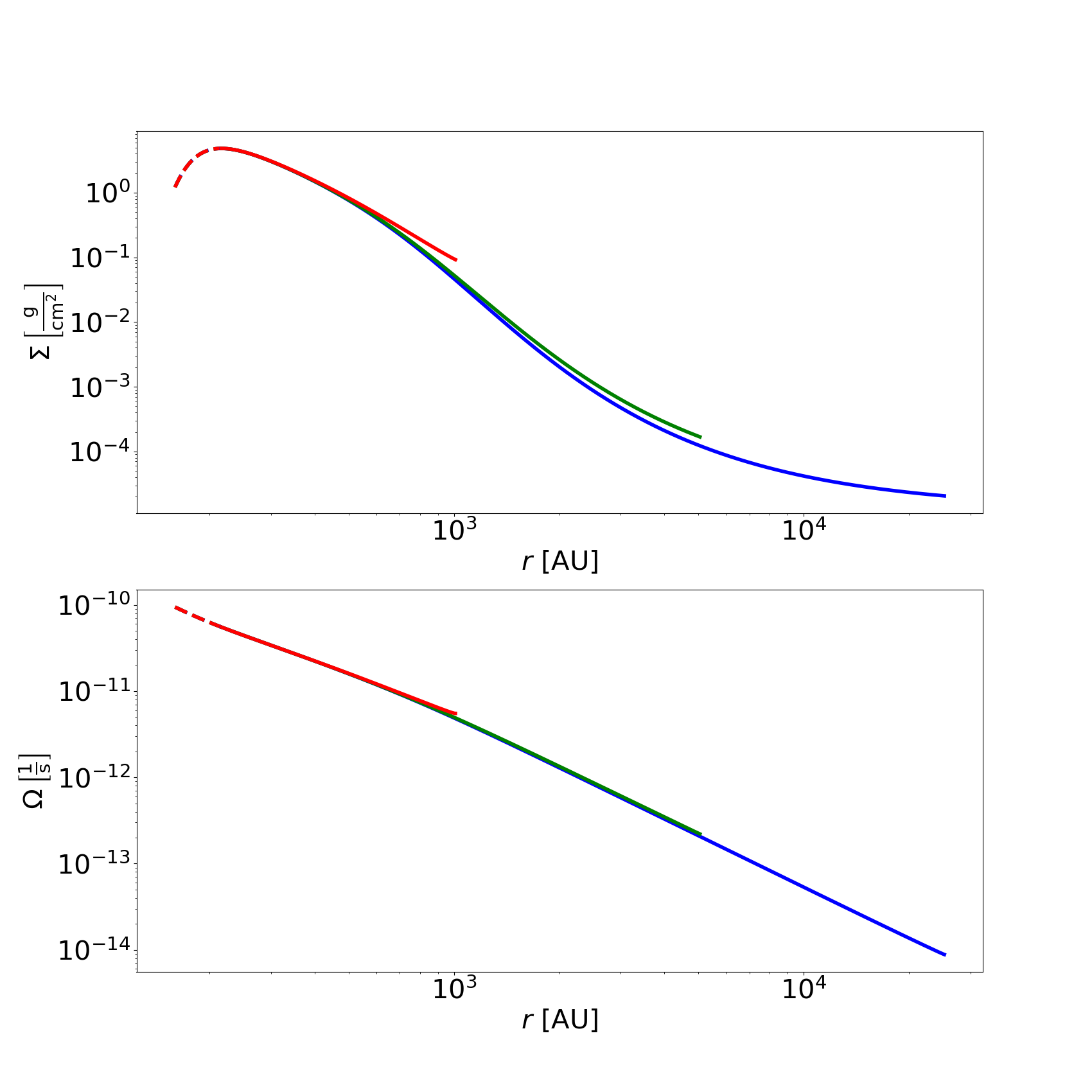}
\includegraphics[width=1\columnwidth]{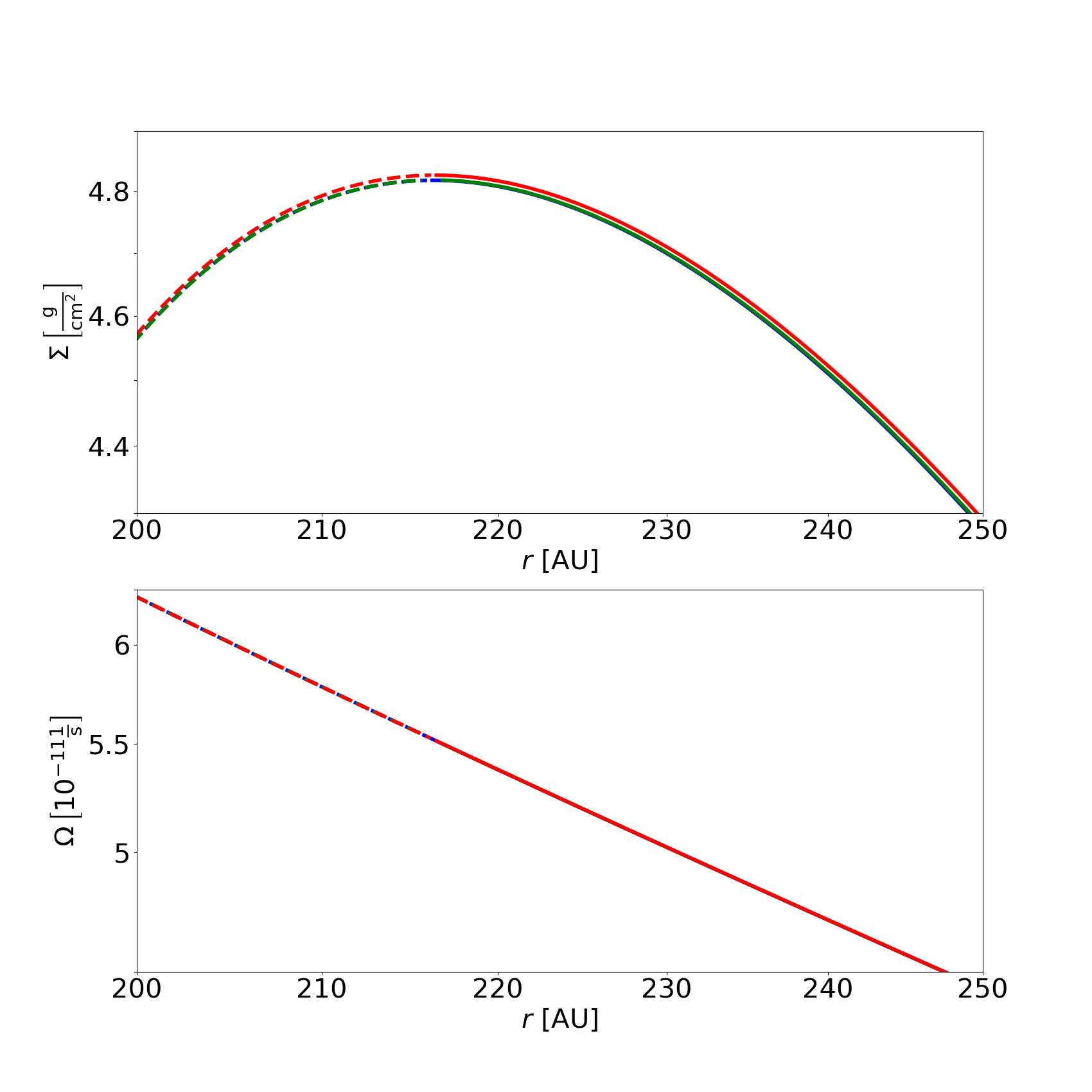}
\caption{Upper panel: Functions $\Sigma(r)$ and $\Omega(r)$ for $\beta=0.03$ (Group C) and $\zeta_{\mathrm{in}}=0.75$ for outer boundaries located at 25,250 AU (blue), 5050 AU (green) and 1010 AU (red). Lower panel: Enlarged images of the same functions around the maximum point of $\Sigma(r)$.}
\label{fig:C-mod}
\end{figure}

The case of $\beta=0.03$ is somewhat different. In this case, there is more mass and angular momentum in the relatively distant parts of the disc, hence, the solution is more sensitive to the displacement of the outer boundary. Figure \ref{fig:C-mod} shows the solutions for $\delta=5$, $\delta=1$ and $\delta=0.2$, i.e., the outer boundary is located at $R_{\mathrm{out}}=25,250 \, \mathrm{AU}$, $R_{\mathrm{out}}=5050 \, \mathrm{AU}$ and $R_{\mathrm{out}}=1010 \, \mathrm{AU}$, respectively. The differences between the solutions are clearly visible. However, it is important to note that these differences vanish as we approach the centre, i.e., the solutions match each other very well in the densest part of the ``outer'' disc, where most of its mass is concentrated.

\section{Comparison with the ``Keplerian'' infall model}
\label{sec:Comparison}

In \cite{Takahashi2013}, the authors supposed that each infalling gas parcel is accreted onto the disc at a radius where the local Keplerian specific angular momentum is equal to that of the gas parcel, and radial pressure gradient plays an insignificant role in establishing a balance of forces. In the following, we examine the conditions under which these assumptions may hold for the outer region of the disc. For this purpose, we calculate the disc structure for different cases using the assumptions of Takahashi et al., omitting, however, the angular momentum transport (the latter is justified by the fact that we are dealing here with the outer part of the disc which is the least affected by this phenomenon during the collapse phase). After that, we check whether the resulting surface density profiles are consistent with the assumption that the role of pressure gradient is negligible, and compare the surface density profiles with those obtained from the calculations presented in this paper.

In the following, we’ll denote by $R_\mathrm{d}$ the disc radius calculated from the assumptions used by Takahashi et al. For a given initial value of $K$ (or $\xi_\mathrm{max}$) and $\beta$, the ratio of $R_\mathrm{d}$ to $R_\mathrm{BE}$ has a unique value which can be calculated as follows:

\begin{equation*}
j = R^2_\mathrm{BE}\omega = \sqrt{G M_{\mathrm{BE}} R_\mathrm{d}} \Rightarrow
\end{equation*}
\begin{equation*}
\begin{split}
R_\mathrm{d} & = \frac{R^4_\mathrm{BE}}{GM_\mathrm{BE}}\omega^2 = \frac{R^4_\mathrm{BE}}{GM_\mathrm{BE}} \cdot \frac{E_\mathrm{BE,rot}}{\kappa M_\mathrm{BE}R^2_\mathrm{BE}} = \frac{R^4_\mathrm{BE}}{GM_\mathrm{BE}} \cdot \frac{\beta |E_\mathrm{BE,grav}|}{\kappa M_\mathrm{BE}R^2_\mathrm{BE}} = \\
& = \frac{R^4_\mathrm{BE}}{GM_\mathrm{BE}} \cdot \frac{\beta}{\kappa M_\mathrm{BE}R^2_\mathrm{BE}} \cdot \lambda \cdot \frac{GM^2_\mathrm{BE}}{R_\mathrm{BE}} = \frac{\beta\lambda}{\kappa} \cdot R_\mathrm{BE} \Rightarrow
\end{split}
\end{equation*}
\begin{equation}
\frac{R_\mathrm{d}}{R_\mathrm{BE}} = \frac{\beta\lambda}{\kappa}
\end{equation}

For a given $\Sigma(r)$ function, one can calculate the ratio $f_\mathrm{p/g}$ between the force arising from pressure gradient and gravity as a function of radial coordinate. If $f_\mathrm{p/g} \ll 1$ holds for most of the mass, then it is justified to neglect the pressure gradient in the model. For a given surface density profile, $f_\mathrm{p/g}$ increases with growing radial coordinate. If one chooses some positive value $A$, one can calculate the mass $M_A$, located beyond a certain radius $R_A$, for which $f_\mathrm{p/g}>A$. The ratio between $M_A$ and $M_\mathrm{BE}$ as well as the ratio between $R_A$ and $R_\mathrm{d}$ depend solely on $K$ and $\beta$.

We have calculated numerically the density profile for several values of $K$ and $\beta$. We produced a Python code which first generates a BE sphere of given parameters, then divides it into a small inner sphere and 2000 spherical shells, slicing the latter to thin rings perpendicularly to the rotation axis. The shells fall successively onto the disc plane in a way that each ring arrives at a radius where the local Keplerian value of specific angular momentum is equal to the specific angular momentum of the ring. (Note that after reaching its place on the disc, each gas parcel remains there, even tough further infall will change the local Keplerian angular velocity.)

Table \ref{table:Kep-infall-model-discs} contains the $\frac{R_A}{R_\mathrm{d}}$ and $\frac{M_A}{M_\mathrm{BE}}$ ratios for $A=0.05$ and $A=0.1$. The values are calculated for discs originating from the collapse of three types of Bonnor-Ebert spheres: a highly subcritical ($K=2$), a slightly supercritical ($K=15$) and a highly supercritical ($K=25$) one with $\beta=0.005$ (for each value of $K$) or $\beta=0.01$, $0.02$ and $0.03$ (only for the supercritical cases).
\begin{figure}[t]
\centering
\includegraphics[width=1\columnwidth]{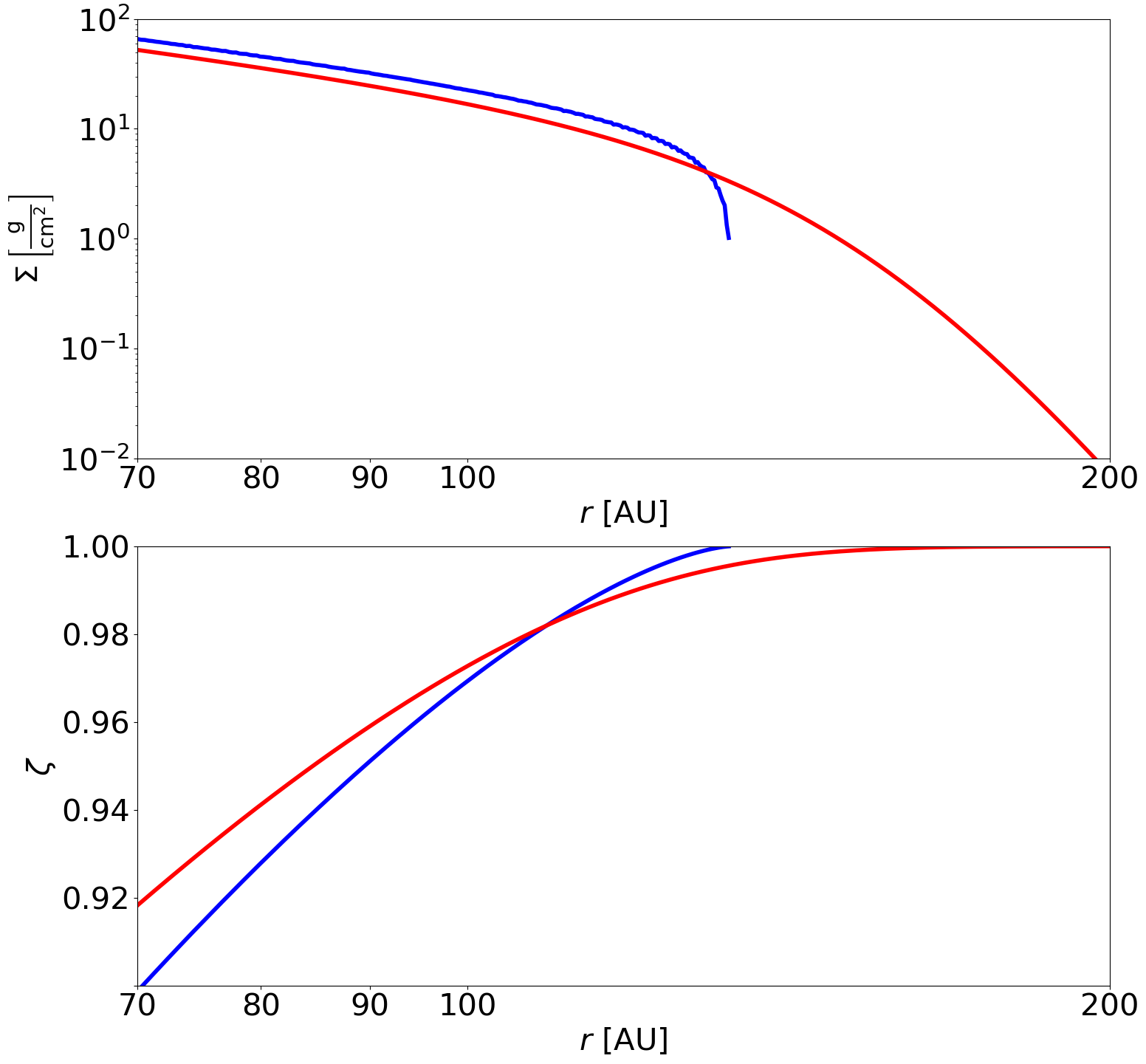}
\includegraphics[width=1\columnwidth]{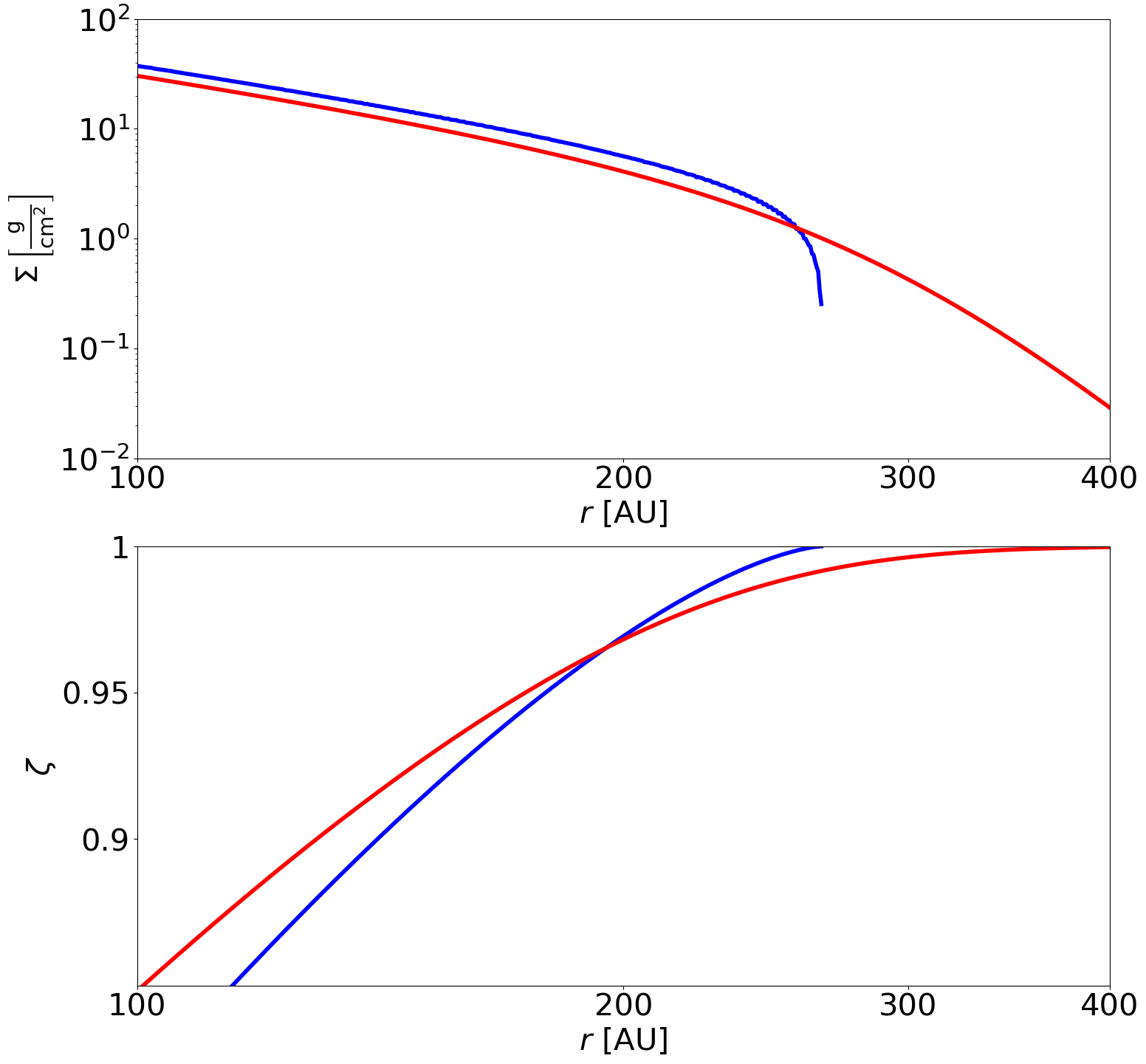}
\caption{$\Sigma(r)$ and $\zeta(r)$ functions for discs arising from the collapse of a slightly supercritical ($K=15$) BE sphere with $\beta=0.005$ (upper panel) and $\beta=0.01$ (lower panel). The red curves correspond to calculations presented in this paper, the blue ones correspond to the ``Keplerian'' infall model.}
\label{fig:comparison00501}
\end{figure}

\begin{figure}[t]
\centering
\includegraphics[width=1\columnwidth]{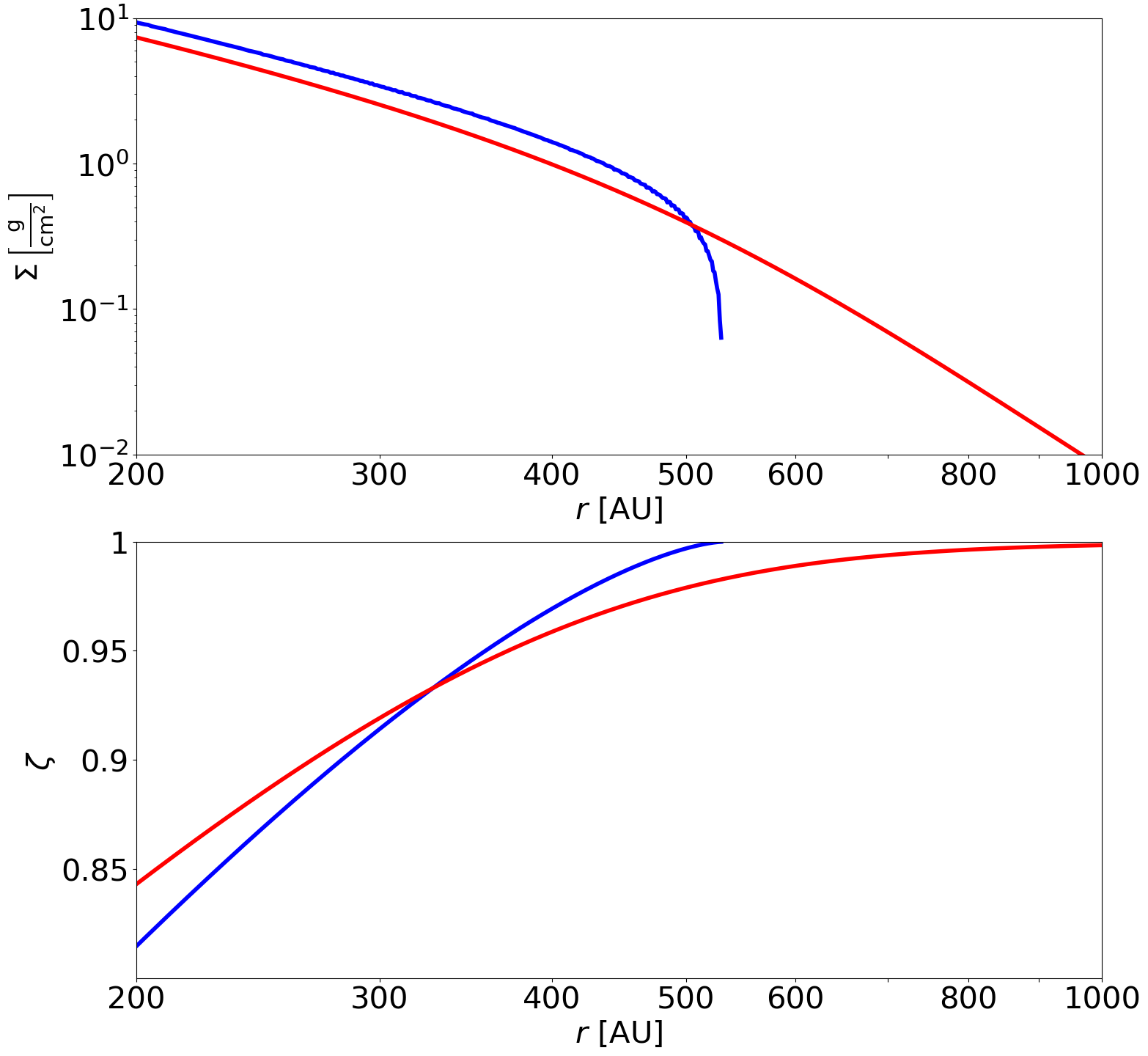}
\includegraphics[width=1\columnwidth]{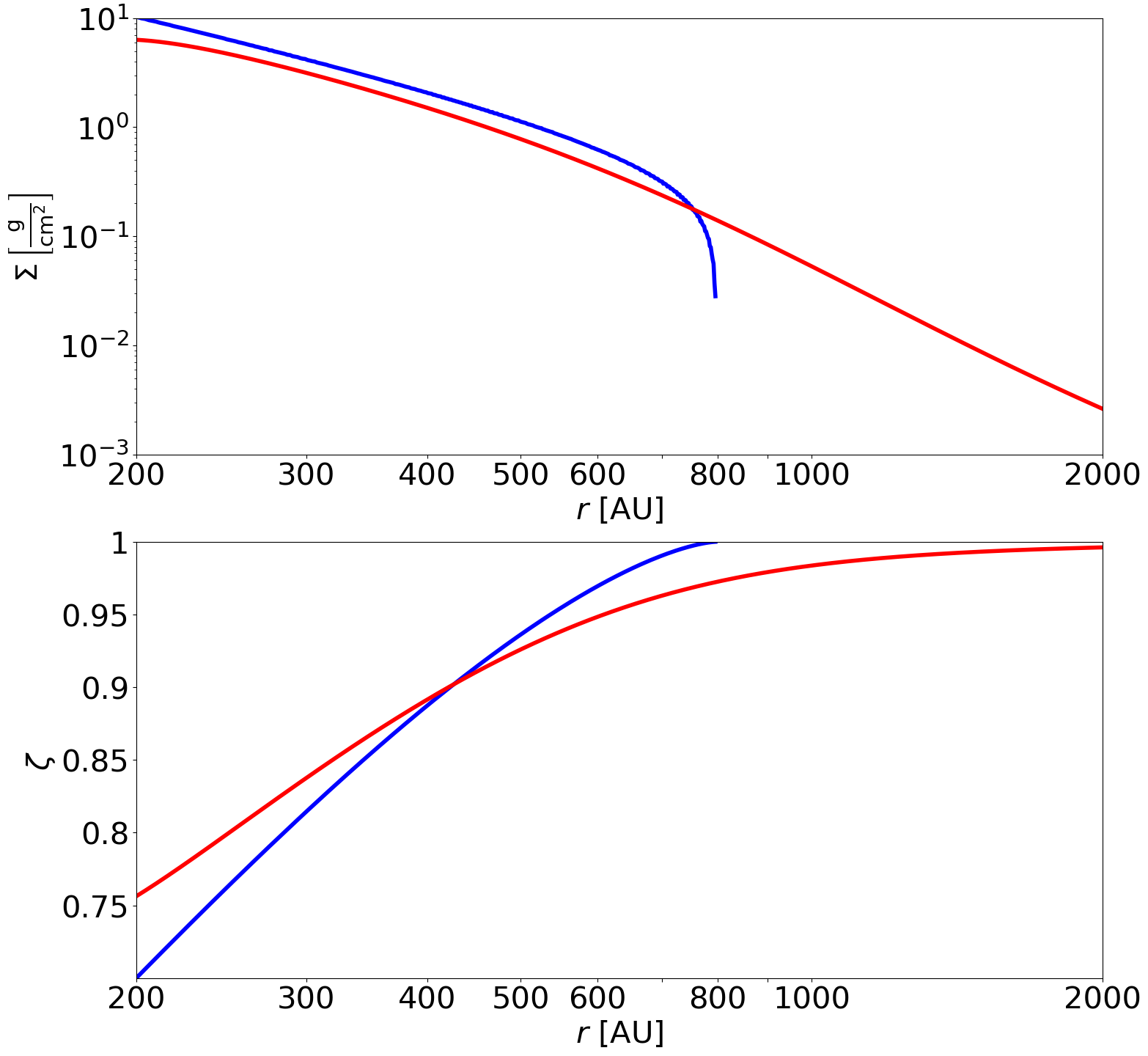}
\caption{$\Sigma(r)$ and $\zeta(r)$ functions for discs arising from the collapse of a slightly supercritical ($K=15$) BE sphere with $\beta=0.02$ (upper panel) and $\beta=0.03$ (lower panel). The red curves correspond to calculations presented in this paper, the blue ones correspond to the ``Keplerian'' infall model.}
\label{fig:comparison0203}
\end{figure}

\begin{table}[ht]
\caption{$R_A$ and $M_A$ ($A=0.05$ and $0.1$) values for some values of $K$ and $\beta$, given in percents of $R_\mathrm{d}$ and $M_{\mathrm{BE}}$, respectively. The absolute values of $R_\mathrm{d}$ and $R_A$ given in the table correspond to a cloud core of 1 solar mass and 20 K temperature.}
\centering
\begin{tabular}{c c c c c c c}
\hline\hline
$\beta$ & $K$ & $R_d$ & $R_{0.05}$ & $M_{0.05}$ & $R_{0.1}$ & $M_{0.1}$ \\
\hline
0.005 & 2 & & 81\% & 6.7\% & 91\% & 1.9\% \\
 & & 186 AU & 150 AU & & 170 AU & \\
\hline
0.005 & 15 & & 86\% & 1.1\% & 94\% & 0.3\% \\
 & & 133 AU & 115 AU & & 125 AU & \\
\hline
0.005 & 25 & & 85\% & 1.2\% & 93\% & 0.3\% \\
 & & 146 AU & 124 AU & & 136 AU & \\
\hline
0.01 & 15 & & 69\% & 4.6\% & 86\% & 1.2\% \\
 & & 266 AU & 183 AU & & 229 AU & \\
\hline
0.01 & 25 & & 66\% & 4.6\% & 85\% & 1.2\% \\
 & & 293 AU & 192 AU & & 248 AU & \\
\hline
0.02 & 15 & & 41\% & 16.6\% & 69\% & 4.6\% \\
 & & 532 AU & 216 AU & & 365 AU & \\
\hline
0.02 & 25 & & 38\% & 15.4\% & 65\% & 4.6\% \\
 & & 586 AU & 221 AU & & 383 AU & \\
\hline
0.03 & 15 & & 25\% & 29.6\% & 53\% & 9.9\% \\
 & & 797 AU & 203 AU & & 425 AU & \\
\hline
0.03 & 25 & & 23\% & 26.8\% & 50\% & 9.6\% \\
 & & 878 AU & 206 AU & & 435 AU & \\
\hline

\end{tabular}
\label{table:Kep-infall-model-discs}
\end{table}

The data from the table suggest the following:

1. For large values of $K$ (i.e., for supercritical BE spheres), $\frac{R_A}{R_\mathrm{d}}$ and $\frac{M_A}{M_\mathrm{BE}}$ ratios weakly depend on $K$. Small values of $K$ (highly subcritical BE spheres) also yield similar $\frac{R_A}{R_\mathrm{d}}$ ratios as large ones, but the corresponding $\frac{M_A}{M_\mathrm{BE}}$ ratios are significantly higher. (However, one may not expect the collapse of highly sub-critical BE spheres.)

2. Both the $\frac{R_A}{R_\mathrm{d}}$ and $\frac{M_A}{M_\mathrm{BE}}$ ratios strongly depend on $\beta$. The observed values of $\beta$ of real cloud cores vary between $10^{-4}$ and 0.07. The role of pressure gradient becomes significant close to the upper end of this interval, i.e., for $\beta\gtrsim0.01$. As we have seen before, the cases with a high degree of self-consistency belong to these values of $\beta$. One shall note, however, that because of rotational flattening, the Bonnor-Ebert sphere becomes a crude approximation of the initial state for large values of $\beta$.

For a better comparison of the disc structures obtained from the two models, we have plotted the corresponding $\Sigma(r)$ and $\zeta(r)$ functions for the four cases with $K=15$ (see Figures \ref{fig:comparison00501} and \ref{fig:comparison0203}).

\section{Summary}

We aimed to determine the structure of the outer part of a disc arising from the collapse of an unstable, non-magnetic Bonnor-Ebert gas sphere initially rotating as a solid body. If certain conservation laws hold (see section 2), it is possible to derive equations for the disc structure without following up the collapse process. In our approach, we may lose the ability to follow planet formation processes that might have already started during the collapse of the molecular cloud as the formation of planetesimals, for instance. On the other hand, \citet{DrazkowskaDullemond2018A&A} have found that planetesimal formation might be uncommon in the disc buildup phase, as requiring specific conditions.

Our model can be valid only for the less dense, ``outer'' part of the disc, therefore our investigations were restricted to this region. The solutions obtained can be fitted to models of the ``inner'' part of the disc, and used this way as initial conditions for the examinations of further disc evolution. (For this end, the solutions can be truncated at an appropriate point, e.g., where $\Omega$ is equal to $\Omega_{\mathrm{Kep}}$, or to a slightly sub-Keplerian value depending on the exponent of power law for the surface density of gas in the inner disc to be fitted to.)

After describing the basic assumptions of the model (section \ref{sec:Description_of_the_physical_model}), we set up the structure equations for the ``outer'' region of the disc in section \ref{sec:Basic_equation_of_structure}. Since these were difficult to handle, in section \ref{sec:Transformed_equations_of_structure} we have transformed them to numerically solvable forms by introducing new functions. Once the transformed equations are integrated, the inverse transformations of the solutions yield us the disc's surface density and angular velocity profiles.

In section \ref{sec:Solutions}, we have examined three groups of cases, corresponding to different angular velocities of the initial Bonnor-Ebert spheres. In the case of the slowest rotation, the resulting solutions are quite sharply divided into three regions: a super-Keplerian, a slighly sub-Keplerian and a highly sub-Keplerian one. (Let us remind that the presence of a super-Keplerian region is a necessary result of arbitrarily choosing the condition expressed in eq. (\ref{eqn:x_in}). It can be considered an artifact and truncated when fitting to a Keplerian ``inner'' disc model.) It is the second region which contains the bulk of the mass. Increasing the angular velocity of the initial BE sphere leads to a fast rise of the resulting disc's size. At the same time, both the width and the mass of the second region diminish, while the third region grows wide and massive.

A comparison with the ``Keplerian'' infall model (which supposes that each gas parcel of the collapsing cloud arrives to the disc at the radius where Keplerian specific angular momentum is equal to that of the gas parcel) in section \ref{sec:Comparison} has shown that for slowly rotating BE spheres, the discs resulting from the two models are very similar, however, there is a considerable difference between the two if the initial BE sphere rotates quickly.

With growing $\beta$, the gravitational stability of the disc increases as well, i.e., more and more mass is located within the stable region of the disc. The collapse of a slowly rotating BE sphere can lead to an early fragmentation, therefore by calculating the Toomre parameter, the gravitationally unstable part of the outer disc can be identified, in which giant planet formation may begin directly from the collapsing fragments.

On the other hand, from the collapse of a quickly rotating BE sphere, an extended, massive and highly sub-Keplerian disc can emerge, which can play the role of a dust reservoir for planet formation after the disc buildup phase.

\begin{acknowledgements}
This research was supported by the NKFIH excellence grant TKP2021-NKTA-64. The authors thank the anonymous reviewer for the useful comments and constructive criticism that helped us considerably improve the manuscript.
\end{acknowledgements}

\bibliographystyle{aa}
\bibliography{biblio}

\begin{appendix}

\section{}

The total mass of the Bonnor-Ebert sphere can be calculated as

\[
M_{\mathrm{BE}}=\int_0^{R_{\mathrm{BE}}} \varrho(r)4\pi r^2 \dd r
\]

Using $\xi=k \cdot r$ and $\varrho(r)=\varrho(0) \cdot F(k \cdot r)$, we get

\[
M_{\mathrm{BE}}=\frac{4\pi\varrho(0)}{k^3} \int_0^{\xi_{\mathrm{max}}} \xi^2 F(\xi) \dd \xi
\]

Using equation (\ref{eqn:k}), we can write for $k^3$ the following:

\[
k^3=k \cdot k^2 = k \cdot \frac{4\pi G \varrho(0) \mu m_{\mathrm{p}}}{k_{\mathrm{B}} T}
\]

The last two equations yield

\[
M_{\mathrm{BE}}=\frac{k_{\mathrm{B}} T}{k G \mu m_{\mathrm{p}}} \int_0^{\xi_{\mathrm{max}}} \xi^2 F(\xi) \dd \xi
\]

From this, we get equation (\ref{eqn:k-2}).

The central density of the Bonnor-Ebert sphere can be written as follows:

\[
\varrho(0)=\frac{M_{\mathrm{BE}}k^3}{4\pi\int_0^{\xi_{\mathrm{max}}}\xi^2 F(\xi) \dd \xi} = \frac{(k_{\mathrm{B}} T)^3}{4\pi(G\mu m_{\mathrm{p}})^3 M_{\mathrm{BE}}^2} \cdot \left( \int_0^{\xi_{\mathrm{max}}}\xi^2 F(\xi) \dd \xi \right)^2
\]

\section{}

Some small part of the gas cloud was initially at a distance $\tilde r$ from the rotation axis. After the collapse, it has a radial coordinate $r$. The conservation of angular momentum yields:

\[
j = \tilde r^2\omega = r^2\Omega \Rightarrow \tilde r = \sqrt{\frac{\Omega}{\omega}} \cdot r
\]

In the initial state, the specific angular momentum interval $[j, j+\dd j]$ corresponds to an interval $[\tilde r, \tilde r+\dd \tilde r]$. The relationship between the widths of these intervals can be written as

\[
\dd j = \frac{\dd j}{\dd \tilde r} \dd \tilde r = 2 \tilde r \omega \dd \tilde r \Rightarrow \dd \tilde r = \frac{\dd j}{2 \tilde r \omega}
\]

The corresponding gas mass $\dd M$:

\begin{equation*}
\begin{split}
\dd M & =2\pi \tilde r \dd \tilde r 2 \cdot \int_0^{\sqrt{R^2-\tilde r^2}} \varrho\left(\sqrt{\tilde r^2+z^2}\right) \dd z = \\
& = \dd j \frac{2\pi}{\omega} \cdot \int_0^{\sqrt{R^2-\frac{j}{\omega}}} \varrho \left(\sqrt{\frac{j}{\omega}+z^2}\right) \dd z
\end{split}
\end{equation*}

Using the notation introduced by equation (\ref{eqn:I_j-function}), we can write:

\[
\dd M =  \dd j \frac{2\pi}{\omega} I(j)
\]

In the final state, the same mass $\dd M$ is distributed within a radial coordinate interval $[r,r+\dd r]$. The width of the interval can be calculated as follows:

\[
\dd j = \frac{\dd j}{\dd r} \dd r = \left(2r\Omega + r^2\frac{\dd \Omega}{\dd r}\right) \dd r \Rightarrow \dd r =\frac{\dd j}{r \cdot \left(2\Omega + r \frac{\dd  \Omega}{\dd r}\right)}
\]

For the mass $\dd M$, we can write

\[
\dd M = 2\pi r \dd r \Sigma = 2\pi\Sigma\frac{\dd j}{2\Omega + r \frac{\dd \Omega}{\dd r}}
\]

The two equations for $\dd M$ yield:

\[
 \dd j \frac{2\pi}{\omega} I(j) =  2\pi\Sigma\frac{\dd j}{2\Omega + r \frac{\dd \Omega}{\dd r}}
\]

Here, the factor $2\pi \dd j$ cancels out. Re-arranging the equation for $\frac{\dd \Omega}{\dd r}$ yields equation (\ref{eqn:dO_dr}).

\section{}

Because of the conservation of angular momentum, the angular velocity of the disc material at $R_{\mathrm{BE}}$ must be the same as the initial angular velocity of the Bonnor-Ebert sphere:

\[
R_{\mathrm{BE}}^2 \cdot \Omega(R_{\mathrm{BE}}) = R_{\mathrm{BE}}^2 \cdot \omega \Rightarrow \Omega(R_{\mathrm{BE}})=\omega
\]

One must suppose that $\Sigma(R_{\mathrm{BE}})$ is not equal to zero, otherwise equation (\ref{eqn:dS_dr}) yields a constant zero solution for $\Sigma(r)$. However, $\Sigma(R_{\mathrm{BE}})>0$ implies that the right hand side of equation (\ref{eqn:dO_dr}) tends to infinity as $r$ approaches to $R_{\mathrm{BE}}$. We assume the following approximate form for $\Omega(r)$ as $r \rightarrow R_{\mathrm{BE}}^-$:

\[
\Omega(r) \cong \omega - c \cdot \Delta r^\frac{1}{n}
\]

Here $c>0$ and $n>1$ are constants, while $\Delta r$ stands for ${R_{\mathrm{BE}}-r}$. The corresponding approximate forms for $j(r)$ and $I(j(r))$ are the following:

\[
j(r) \cong R_{\mathrm{BE}}^2 \cdot \left( \omega - c \cdot \Delta r^\frac{1}{n} \right) = j_{\mathrm{max}} - R_{\mathrm{BE}}^2 \cdot c \cdot \Delta r^\frac{1}{n}
\]

\[
I(j(r)) \cong \sqrt{R_{\mathrm{BE}}^2-\frac{j(r)}{\omega}} \cdot \varrho(R_{\mathrm{BE}}) \cong \sqrt{c} \cdot R_{\mathrm{BE}} \cdot \varrho(R_{\mathrm{BE}}) \cdot \Delta r^\frac{1}{2n}
\]

For the two sides of equation (\ref{eqn:dO_dr}) we have:

\[
\frac{\dd \Omega}{\dd r} \propto \Delta r^{\frac{1}{n}-1}
\] 

\[
\frac{\omega}{r \cdot I(j(r))} \cdot \Sigma - \frac{2\Omega}{r} \propto \frac{1}{I(j(r))} \propto \Delta r^{-\frac{1}{2n}}
\]

The two exponents must be equal, which yields the value of $n$:

\[
\frac{1}{n}-1=-\frac{1}{2n} \Rightarrow n=\frac{3}{2}
\]

From these results it follows that $1-\frac{j(r)}{j_{\mathrm{max}}} \propto \Delta r^\frac{2}{3}$. Therefore we introduce the function $\left(1-\frac{j(r)}{j_{\mathrm{max}}}\right)^{\frac{3}{2}}$, which -- unlike $\Omega(r)$ -- has a finite (but non-zero) derivative at $R_{\mathrm{BE}}$.

\section{}

In subsection (\ref{grav_pot_gradient}) we introduced the function $\zeta(\chi)$. For any $\chi'\in[0,1]$, $\zeta(\chi')$ is equal to the sum of all mass with $\chi>\chi'$, normalized by total mass of the cloud core. Below we derive the expression for $\zeta(\chi)$ given by eq. (\ref{def:zeta}).

According to eq. (\ref{def:chi}), $\chi>\chi'$ is equivalent to
\begin{equation*}
j<j'=\left(1-\chi'^{\frac{2}{3}}\right)\cdot j_{\mathrm{max}}
\end{equation*}

Since the cloud core is supposed to rotate initially as a solid body, one can calculate the mass $M'$ with $j<j'$ as the integral of density over the intersection of the Bonnor-Ebert sphere and an infinite cylinder with radius $r'=\sqrt{j'/\omega}$:
\begin{equation*}
\begin{split}
M' & =\int_0^{\sqrt{j'/\omega}} \left(\int_0^{\sqrt{R_{\mathrm{BE}}^2-r^2}} \varrho\left(\sqrt{r^2+z^2}\right) \dd z \right)2\pi r\dd r = \\
& =\int_0^{\sqrt{j'/\omega}} \left(\int_0^{R_{\mathrm{BE}}\sqrt{1-r^2/R_{\mathrm{BE}}^2}} \varrho\left(\sqrt{r^2+z^2}\right) \dd z \right)2\pi r\dd r
\end{split}
\end{equation*}

Instead of $r$ and $z$, we can introduce the variables $\varepsilon=\frac{r^2}{R^2_{\mathrm{BE}}}=\frac{j}{j_{\mathrm{max}}}=1-\chi^{\frac{2}{3}}$ and $z'=\frac{z}{R_{\mathrm{BE}}}$. Then we have
\begin{equation*}
M'=\int_0^{1-\chi'^{\frac{2}{3}}} \left(\int_0^{\sqrt{1-\varepsilon}} \varrho\left(\sqrt{\varepsilon R^2_{\mathrm{BE}}+z'^2 R^2_{\mathrm{BE}}}\right) R_{\mathrm{BE}} \dd z' \right)\pi R^2_{\mathrm{BE}}\dd \varepsilon
\end{equation*}

Since $\varrho(r)=\varrho_0\cdot F(k\cdot r)$ and $k\cdot R_{BE}=\xi_{max}$,
\begin{equation*}
\begin{split}
& \varrho\left(\sqrt{\varepsilon R^2_{\mathrm{BE}}+z'^2 R^2_{\mathrm{BE}}}\right)=\varrho\left(R_{\mathrm{BE}}\sqrt{\varepsilon +z'^2}\right)=\\
& =\varrho_0\cdot F\left(k\cdot R_{\mathrm{BE}}\sqrt{\varepsilon +z'^2}\right)=\varrho_0\cdot F\left(\xi_{\mathrm{max}}\sqrt{\varepsilon +z'^2}\right)
\end{split}
\end{equation*}

Thus,
\begin{equation*}
M'=\pi\varrho_0 R^3_{\mathrm{BE}}\int_0^{1-\chi'^{\frac{2}{3}}} \left(\int_0^{\sqrt{1-\varepsilon}} F\left(\xi_{\mathrm{max}}\sqrt{\varepsilon+z'^2}\right) \dd z' \right)\dd \varepsilon
\end{equation*}

By substituting $\chi'=0$, we can get from this formula the whole mass of the Bonnor-Ebert sphere:
\begin{equation*}
M_{\mathrm{BE}}=\pi\varrho_0 R^3_{\mathrm{BE}}\int_0^{1} \left(\int_0^{\sqrt{1-\varepsilon}} F\left(\xi_{\mathrm{max}}\sqrt{\varepsilon+z'^2}\right) \dd z' \right)\dd \varepsilon
\end{equation*}

$\zeta(\chi')$ is equal to $\frac{M'}{M_\mathrm{BE}}$. By dividing the last two expressions above, we get eq. (\ref{def:zeta}).

\section{}

The parameter $\kappa$ denotes the ratio of the Bonnor-Ebert sphere's rotational energy $E_{\mathrm{BE,rot}}$ to the product $M_{\mathrm{BE}}R_{\mathrm{BE}}^2\omega^2$. Its value is determined by the dimensionless radius $\xi_{\mathrm{max}}$ of the sphere. To calculate it, at first, we introduce the Bonnor-Ebert sphere's (scalar) moment of inertia $I_{\mathrm{BE}}$. The value of $I_{\mathrm{BE}}$ can be calculated as the sum of contributions from thin spherical shells, each of which has a uniform mass distribution:
\begin{equation*}
I_{\mathrm{BE}} = \int_0^{R_{\mathrm{BE}}} \frac{2}{3} r^2 \varrho(r) 4\pi r^2 \dd r
\end{equation*}

By switching variables, one can transform this equation to
\begin{equation*}
I_{\mathrm{BE}} = \frac{8\pi\varrho(0)}{3k^5} \int_0^{\xi_{\mathrm{max}}} \xi^4 F(\xi) \dd \xi
\end{equation*}

Thus, rotational energy can be written as
\begin{equation*}
E_{\mathrm{BE,rot}} = \frac{1}{2} I_{\mathrm{BE}} \omega^2 = \frac{4\pi\varrho(0)}{3k^5} \omega^2 \int_0^{\xi_{\mathrm{max}}} \xi^4 F(\xi) \dd \xi
\end{equation*}

Since $\varrho(0)=\frac{M_{\mathrm{BE}}k^3}{4\pi\int_0^{\xi_{\mathrm{max}}}\xi^2 F(\xi) \dd \xi}$ (cf. Appendix A) and $k=\frac{\xi_{\mathrm{max}}}{R_{\mathrm{BE}}}$ (see equation (\ref{eqn:k-def})), for the factor $\frac{4\pi \varrho(0)}{3k^5}$ we have:
\begin{equation*}
\frac{4\pi\varrho(0)}{3k^5} = \frac{4\pi}{3k^5} \cdot \frac{M_{\mathrm{BE}}k^3}{4\pi\int_0^{\xi_{\mathrm{max}}}\xi^2 F(\xi) \dd \xi} = \frac{M_{\mathrm{BE}}R_{\mathrm{BE}}^2}{3\xi_{\mathrm{max}}^2 \cdot \int_0^{\xi_{\mathrm{max}}}\xi^2 F(\xi) \dd \xi}
\end{equation*}

Therefore,
\begin{equation*}
E_{\mathrm{BE,rot}} = \frac{M_{\mathrm{BE}}R_{\mathrm{BE}}^2 \omega^2}{3\xi_{\mathrm{max}}^2 \cdot \int_0^{\xi_{\mathrm{max}}}\xi^2 F(\xi) \dd \xi} \int_0^{\xi_{\mathrm{max}}} \xi^4 F(\xi) \dd \xi
\end{equation*}
\begin{equation*}
\kappa = \frac{E_{\mathrm{BE,rot}}}{M_{\mathrm{BE}}R_{\mathrm{BE}}^2 \omega^2} = \frac{\int_0^{\xi_{\mathrm{max}}} \xi^4 F(\xi) \dd \xi}{3\xi_{\mathrm{max}}^2 \cdot \int_0^{\xi_{\mathrm{max}}}\xi^2 F(\xi) \dd \xi}
\end{equation*}

By $\lambda$ we have denoted the ratio of the absolute value of gravitational energy to the fraction $\frac{GM_{\mathrm{BE}}^2}{R_{\mathrm{BE}}}$. The absolute value of the Bonnor-Ebert sphere's total gravitational energy can be calculated as follows:
\begin{equation*}
\begin{split}
|E_{\mathrm{BE,grav}}| & = \int_0^{R_{\mathrm{BE}}} \frac{G \int_0^r \varrho(r') 4\pi r'^2 \dd r'}{r} \varrho(r) 4\pi r^2 \dd r =\\
& = \frac{16\pi^2 G \varrho^2(0)}{k^5} \int_0^{\xi_{\mathrm{max}}} \left( \int_0^{\xi} \xi'^2 F(\xi') \dd \xi' \right) \xi F(\xi) \dd \xi =\\
&= \frac{\xi_{\mathrm{max}} \cdot \int_0^{\xi_{\mathrm{max}}} \left( \int_0^{\xi} \xi'^2 F(\xi') \dd \xi' \right) \xi F(\xi) \dd \xi}{\int_0^{\xi_{\mathrm{max}}}\xi^2 F(\xi) \dd \xi} \cdot \frac{GM_{\mathrm{BE}}^2}{R_{\mathrm{BE}}}
\end{split}
\end{equation*}

Thus, $\lambda$ is given by the formula

\begin{equation*}
\lambda = \frac{\xi_{\mathrm{max}} \cdot \int_0^{\xi_{\mathrm{max}}} \left( \int_0^{\xi} \xi'^2 F(\xi') \dd \xi' \right) \xi F(\xi) \dd \xi}{\int_0^{\xi_{\mathrm{max}}}\xi^2 F(\xi) \dd \xi}
\end{equation*}

We have defined a third dimensionless parameter, $\tau$, which denotes the ratio of the absolute value of gravitational energy to the product $\lambda M_{\mathrm{BE}} c_{\mathrm{s}}^2$. The isothermal sound speed $c_{\mathrm{s}}$ is equal to $\sqrt{\frac{k_{\mathrm{B}} T}{\mu m_{\mathrm{p}}}}$. Using equations ($\ref{eqn:k-def}$) and (\ref{eqn:k-2}), we have:
\begin{equation*}
M_{\mathrm{BE}} c_{\mathrm{s}}^2 = M_{\mathrm{BE}} \frac{k_{\mathrm{B}} T}{\mu m_{\mathrm{p}}} =  M_{\mathrm{BE}} k \frac{G M_{\mathrm{BE}}}{\int_0^{\xi_{\mathrm{max}}} \xi^2 F(\xi) \dd \xi} = \frac{\xi_{\mathrm{max}}}{\int_0^{\xi_{\mathrm{max}}} \xi^2 F(\xi) \dd \xi} \cdot \frac{G M_{\mathrm{BE}}^2}{R_{\mathrm{BE}}}
\end{equation*}

Therefore,
\begin{equation*}
\tau = \frac{|E_{\mathrm{BE,grav}}|}{\lambda M_{\mathrm{BE}} c_{\mathrm{s}}^2} = \frac{\lambda \frac{G M_{\mathrm{BE}}^2}{R_{\mathrm{BE}}}}{\lambda \frac{\xi_{\mathrm{max}}}{\int_0^{\xi_{\mathrm{max}}} \xi^2 F(\xi) \dd \xi} \cdot \frac{G M_{\mathrm{BE}}^2}{R_{\mathrm{BE}}}} = \frac{\int_0^{\xi_{\mathrm{max}}} \xi^2 F(\xi) \dd \xi}{\xi_{\mathrm{max}}}
\end{equation*}

In order to derive equation (\ref{eqn:ds_dx-final}), we depart from equations (\ref{eqn:ds_dx}) and (\ref{eqn:dV_dx}), making the following transformations:
\begin{equation*}
\begin{split}
\frac{\dd \sigma}{\dd x} & = - \frac{R_{\mathrm{BE}}^2\omega^2}{c_{\mathrm{s}}^2}\frac{\left(1-\chi^{2/3}\right)^2}{(1-x)^3} - \frac{1}{c_{\mathrm{s}}^2} \left( - \frac{GM_{\mathrm{BE}}\zeta(\chi)}{R_{\mathrm{BE}}(1-x)^2} \right) = \\
& = \frac{1}{M_{\mathrm{BE}}c_{\mathrm{s}}^2} \left(  \frac{GM_{\mathrm{BE}}^2\zeta(\chi)}{R_{\mathrm{BE}}(1-x)^2} - M_{\mathrm{BE}} R_{\mathrm{BE}}^2\omega^2\frac{\left(1-\chi^{2/3}\right)^2}{(1-x)^3} \right) = \\
& = \frac{1}{M_{\mathrm{BE}}c_{\mathrm{s}}^2} \left(  \frac{|E_{\mathrm{BE,grav}}|\zeta(\chi)}{\lambda (1-x)^2} - E_{\mathrm{BE,rot}}\frac{\left(1-\chi^{2/3}\right)^2}{\kappa (1-x)^3} \right) = \\
& = \frac{1}{\lambda M_{\mathrm{BE}}c_{\mathrm{s}}^2} \left(  \frac{|E_{\mathrm{BE,grav}}|\zeta(\chi)}{(1-x)^2} - \beta |E_{\mathrm{BE,grav}}|\frac{\lambda \left(1-\chi^{2/3}\right)^2}{\kappa (1-x)^3} \right) = \\
& = \frac{|E_{\mathrm{BE,grav}}|}{\lambda M_{\mathrm{BE}}c_{\mathrm{s}}^2} \left( \frac{\zeta(\chi)}{(1-x)^2} - \frac{\beta \lambda \left(1-\chi^{2/3}\right)^2}{\kappa (1-x)^3} \right) = \\
& = \tau \left( \frac{\zeta(\chi)}{(1-x)^2} - \frac{\beta \lambda \left(1-\chi^{2/3}\right)^2}{\kappa (1-x)^3} \right)
\end{split}
\end{equation*}
\end{appendix}
\end{document}